\documentclass[nonacm]{acmart}

\AtBeginDocument{%
  \providecommand\BibTeX{{%
    \normalfont B\kern-0.5em{\scshape i\kern-0.25em b}\kern-0.8em\TeX}}}

\setcopyright{rightsretained}
\acmDOI{10.1145/1122445.1122456}

\usepackage[colorinlistoftodos]{todonotes}
\usepackage{xspace}

\usepackage[normalem]{ulem}

\usepackage{enumitem}
\usepackage{tabularx}
\newcolumntype{L}[1]{>{\raggedright\let\newline\\\arraybackslash\hspace{0pt}}p{#1}}
\newcolumntype{M}[1]{>{\raggedright\let\newline\\\arraybackslash\hspace{0pt}}m{#1}}

\newcommand{\systemName}{\textsc{Paper Plain}\xspace}

\newcommand{\reader}{Sarah\xspace}

\newcommand{\keyQs}{Key Question Index and Answer Gists\xspace}
\newcommand{\plainSum}{Section Gists\xspace}
\newcommand{\defs}{Term Definitions\xspace}

\newcommand{\navGuide}{Questions and Answers\xspace}
\newcommand{\inSituExpl}{Sections and Terms\xspace}
\newcommand{\pdf}{PDF baseline\xspace}

\newcommand{\ta}[1]{}
\newcommand{\llw}[1]{}
\newcommand{\kl}[1]{}
\newcommand{\ah}[1]{}
\newcommand{\jb}[1]{}
\newcommand{\mh}[1]{}
\newcommand{\dw}[1]{}

\begin{document}

\title[\systemName: Making Medical Research Papers Approachable to Healthcare Consumers with Natural Language Processing]{\systemName: Making Medical Research Papers Approachable to Healthcare Consumers with Natural Language Processing}
 

\author{Tal August}
\email{taugust@cs.washington.edu}
\affiliation{%
  \institution{University of Washington}
  \city{Seattle}
  \state{Washington}
  \country{USA}
}

\author{Lucy Lu Wang}
\email{lucyw@allenai.org}
\affiliation{%
  \institution{Allen Institute for Artificial Intelligence}
  \city{Seattle}
  \state{Washington}
  \country{USA}
}

\author{Jonathan Bragg}
\email{jbragg@allenai.org}
\affiliation{%
  \institution{Allen Institute for Artificial Intelligence}
  \city{Seattle}
  \state{Washington}
  \country{USA}
}

\author{Marti A. Hearst}
\email{hearst@berkeley.edu}
\affiliation{%
  \institution{University of California, Berkeley}
  \city{Berkeley}
  \state{California}
  \country{USA}
}

\author{Andrew Head}
\email{head@seas.upenn.edu}
\affiliation{%
  \institution{University of Pennsylvania}
  \city{Philadelphia}
  \state{Pennsylvania}
  \country{USA}
}

\author{Kyle Lo}
\email{kylel@allenai.org}
\affiliation{%
  \institution{Allen Institute for Artificial Intelligence}
  \city{Seattle}
  \state{Washington}
  \country{USA}
}

\renewcommand{\shortauthors}{}

\begin{abstract}

When seeking information not covered in patient-friendly documents, like medical pamphlets, healthcare consumers may turn to the research literature. Reading medical papers, however, can be a challenging experience. To improve access to medical papers, we introduce a novel interactive interface---\systemName{}\footnote{Code and demo available at \url{https://github.com/allenai/paper-plain}}---with four features powered by natural language processing: definitions of unfamiliar terms, in-situ plain language section summaries, a collection of key questions that guide readers to answering passages, and plain language summaries of the answering passages. We evaluate \systemName{}, finding that participants who use \systemName{} have an easier time reading and understanding research papers without a loss in paper comprehension compared to those who use a typical PDF reader. Altogether, the study results suggest that guiding readers to relevant passages and providing plain language summaries, or ``gists,'' alongside the original paper content can make reading medical papers easier and give readers more confidence to approach these papers.

\end{abstract}




\maketitle

\section{Introduction}

\begin{figure}
\includegraphics[width=\textwidth]{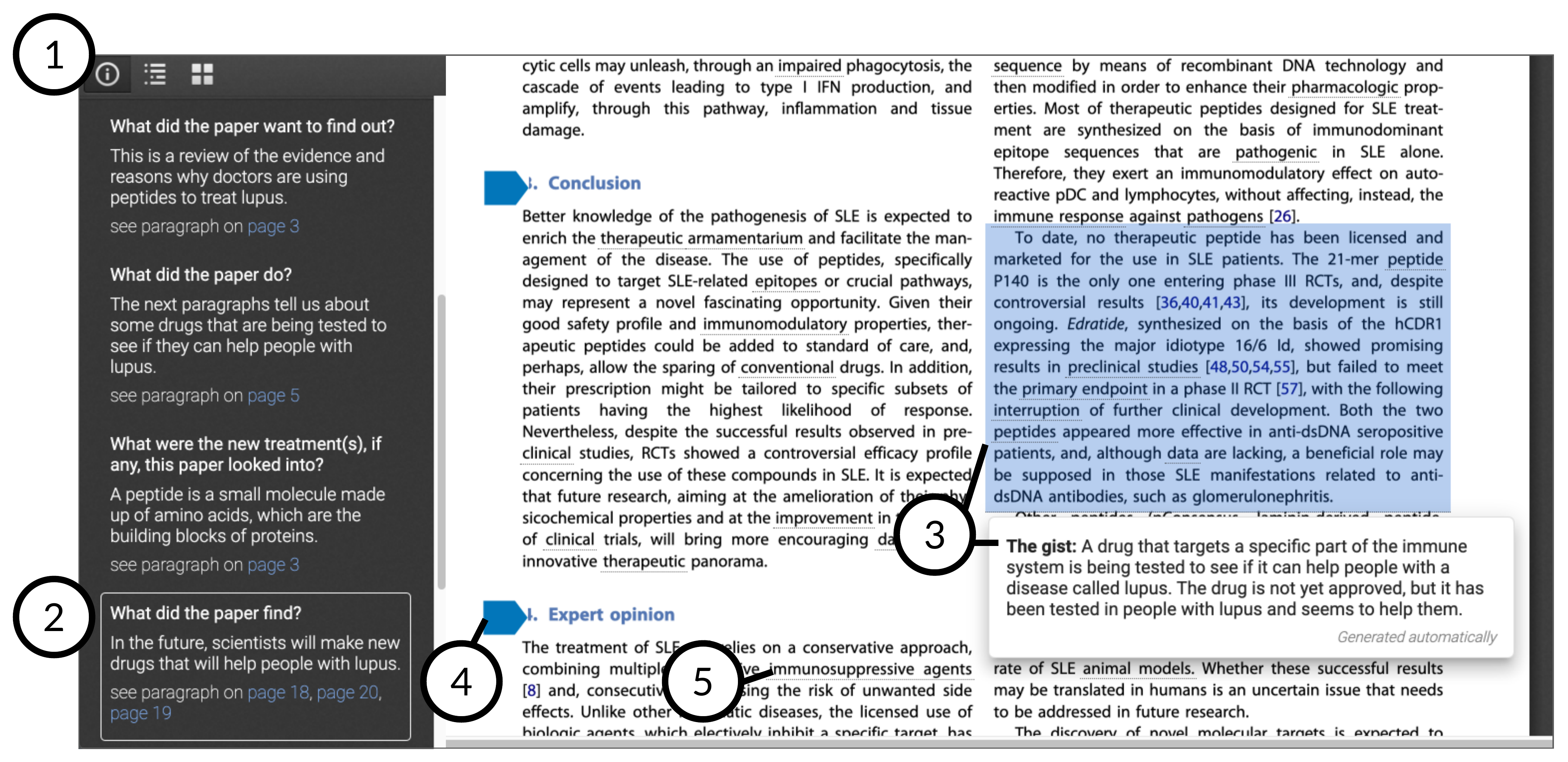}
\caption{\systemName{} helps healthcare consumers consult medical research papers by making their texts more approachable. Shown is the \systemName{} interface and the assistive features it provides to readers. When a paper is opened in \systemName{}, a side pane opens with a reading guide (1), comprising of curated key questions a reader might ask, previews of generated plain language answers, and pointers to where in the paper the reader can find more details. When a reader clicks a question (2), the paper jumps to the passage that provides that answer, accompanied by the plain language answer (answer gist) (3). Readers can also access in-situ plain language summaries for every section of the paper (section gists) by clicking labels next to section headers (4), and definitions of medical jargon sourced from external references by clicking those terms (5).}
\label{fig:teaser}
\end{figure}

A strong public health system depends on the timely dissemination of medical findings to those who need them. Most often, healthcare consumers stay apprised of medical findings through communication with experts---conversation with their doctors, printed materials like pamphlets, and online resources like MedlinePlus or hospital websites~\cite{Zhang2014SearchingFS, Kivits2006InformedPA, Cocco2018DrGI}. However, these resources do not cover all medical conditions and treatments \cite{Rogers2017InternetDeliveredHI, Baskin2020GapsIO}, especially those which are the focus of emerging research \cite{Rabeharisoa2013EvidencebasedAP, Brown2004EmbodiedHM}. In these cases, healthcare consumers may have no choice but to go to the source of medical knowledge---the research literature~\cite{Zuccala2010OpenAA, nationalpolicy, Tennant2016TheAE, Day2020OpenTT, Epstein1997ImpureSA}. In the words of one patient:\footnote{Example from patient testimonials on ``Who needs access?'' an open-access advocacy site, \url{https://whoneedsaccess.org/category/patients/}, content furnished under the Creative Commons Attribution 3.0 Unported License (CC BY 3.0).}

\begin{quote}
\textit{%
 I had been studying CLL [Chronic Lymphocytic Leukemia] through free access articles on PubMed and Google Scholar\ldots{} Reading these NIH papers enabled me to have an intelligent dialogue with a CLL specialist, ultimately leading me to the selection of a clinical trial.
}
\end{quote}


However, it is one thing for healthcare consumers to access the literature, and another thing entirely for them to comfortably navigate research papers. Healthcare consumers report that, unsurprisingly, medical papers are difficult to read~\cite{Day2020OpenTT, Nunn2014LaySO}. This is in part due to being overwhelmed by the amount of unfamiliar jargon. It is also because healthcare consumers are unaccustomed to the norms of how research is conducted and how reports of it are structured~\cite{Britt2014ScientificLT, Day2020OpenTT}. The result is that reading medical papers can be an experience that is challenging and at times demoralizing.

Given these difficulties, is it helpful and effective for healthcare consumers to read medical research papers? We believe that interacting with these papers gives patients an awareness of cutting edge medical findings and the complexities of underlying studies, even if they do not fully comprehend them. These papers also constitute the literature patients wish to share with their healthcare providers, should they discover information germane to treatment options~\cite{Day2020OpenTT, Zuccala2010OpenAA, nationalpolicy}.


In this paper, we ask how interactive information interfaces can make the research literature approachable to non-expert healthcare consumers that need it---whom we refer to as ``readers'' in this paper. In particular, we study how the paper itself can be imbued with new affordances to help readers navigate and evaluate its contents. As the human-computer interaction literature shows, reading interfaces can offer novel affordances to assist readers in navigating documents in new ways~\cite{Abekawa2016SideNoterSP}, looking up the meaning of unfamiliar terminology~\cite{Head2021AugmentingSP}, summarizing sections as they read~\cite{Bohn2021HoneAY}, and searching for answers to their questions~\cite{Zhao2020TalkTP}. Drawing on this work as inspiration, we ask what combination of affordances would be necessary to help bridge the often enormous gap between a reader's current knowledge of biomedical research and the contents of a paper. Consider, for instance, this sentence from a recent article about systemic lupus erythematosus, linked to from a patient-facing MedlinePlus page~\cite{Talotta2020TherapeuticPF}: 


\begin{quote}
\emph{%
The most salient events include an impaired apoptosis of dying cells, a type I interferon (IFN) signature, the uncontrolled activation of T and B lymphocytes and the production of autoantibodies mainly directed against nucleic acids or ribonucleoproteins (RNP).
}
\end{quote}

This sentence is difficult not only because it contains technical jargon, but that \emph{in combination} these words form a sentence so foreign that a reader has little chance of understanding it without learning a considerable amount of background knowledge from elsewhere. A reader not only needs to know what ``autoantibodies'' and ``ribonucleoproteins'' mean, but also how production of one implies condition progression and risks to their health. A medical paper contains not one but hundreds of such sentences, making it exceedingly difficult for readers to find, let alone understand, information important to them. We believe that future interactive aids will need to go beyond their typical capabilities to instead help readers understand where to find information of interest in a paper according to the language they already know.


The key insight of this paper is that medical papers can be made more approachable by judiciously incorporating plain language summaries to supplement original paper content. A reader can engage with the original text through plain language summaries---which we refer to as ``gists''---that contain simplified sentences and reduced jargon and are presented alongside passages in the paper. The reader can approach any content in the paper by first inspecting its gist, only committing attention to a dense passage after learning if it is likely to be relevant. In this way, the reader has the support to engage meaningfully with the original paper text: skipping passages of little relevance and spending time reading those of consequence.

This paper begins with a formative observational study of 12 non-expert readers to identify barriers in reading medical research papers. We observed that, in addition to the expected pervasive difficulties of understanding passages dense with unknown terminology, readers struggled to know what parts of a paper to read and often spent considerable effort making sense of sections with limited usefulness to them. These findings suggest that reading medical papers is uniquely challenging for our envisioned readers due to their lack of domain knowledge and understanding of how medical research is communicated. An augmented reading interface for these readers will need to go beyond the capabilities of prior interfaces---that define terminology~\cite{Head2021AugmentingSP}, provide summaries~\cite{Guo2021AutomatedLL}, or allow readers to ask questions of a paper~\cite{Zhao2020TalkTP}---and provide a reading experience that guides readers to useful information and helps them understand this information in the context of the paper.



To improve access to medical papers, we develop a novel interactive system, \systemName{}, through an iterative design process. The system is designed to make medical papers accessible with four features (illustrated in Figure~\ref{fig:teaser}) that combine to provide support at multiple levels of granularity (e.g., term, paragraph, section) and throughout the reading process. \systemName{} helps a reader find information relevant to them in the paper by providing a list of key questions about medical studies and a preview of plain language answers (``key question index''). When a question is clicked, it takes readers to the paragraphs that answer the question along with full paired plain language answers (``answer gists''). \systemName{} helps a reader understand the essence of jargon-dense passages by allowing them to access in-situ plain language summaries of any section (``section gists''). Finally, \systemName{} provides assistance for understanding unfamiliar terms by making their definitions available. The first three of these features (``key question index'', ``answer gists'' and ``section gists'') are novel in the context of reading applications for research papers, while the fourth is a known feature, though it is necessary to provide holistic reading support. The design of the system is described at length in \S\ref{sec:design}.

We envision \systemName{} as a system that can one day be enabled for any medical research paper. The system draws on active research in natural language processing for biomedical question answering~\cite{10.1007/978-3-030-43887-6_64}, plain language generation~\cite{Guo2021AutomatedLL}, and term identification~\cite{neumann-etal-2019-scispacy}. One limitation of current text generation capablities is the risk of generating factually incorrect or inconsistent text, often referred to as ``hallucinations''~\cite{Maynez2020OnFA}. Deploying any system in a medical context will require algorithmic advances or human oversight to detect factually incorrect generations~\cite{Kryscinski2020EvaluatingTF, Maynez2020OnFA}. In this project we assume such advances are possible (see~\cite{Gabriel2021DiscourseUA, Laban2021SummaCRN} for examples of current advances in this space) and provide some manual filtering of incorrect, incoherent, or copied text (i.e., selecting the most fluent and correct generation out of five). This allows us to focus on developing interactions that would enable readers to meaningfully engage with medical research papers. \S\ref{sec:implementation} describes the implementation of \systemName{} and highlights the adaptations needed to make text generation model outputs useful for readers, while \S\ref{sec:ethicalImplications} discusses in more depth the limitations of text generation models for our application. While to date our implementation relies on some human curation, this project as a whole indicates the potential for reading experiences like \systemName{} to be deployed at scale over the scientific literature.

To assess how \systemName{} supports the reading experience, we conducted a 24 within-participant usability study where participants read papers with variants of \systemName{} or a typical PDF reader during a timed reading task. The study showed that \systemName{} lowered participants' self-reported difficulty in reading the paper and increased confidence that they found all of the information of interest to themselves without any observable degradation in paper comprehension. The clear favorite feature was the key question index and answer gists. Participants also used, and appreciated, in-situ section gists and term definitions; though participants tended not to use them when the aforementioned key question-based navigation was available. Altogether, this study suggests that reading interfaces that provide guidance and plain language summaries can make medical papers more approachable and offer readers more confidence than they would otherwise have when reading medical research papers. 

In summary, this paper contributes:

\begin{enumerate}
    \item A characterization of the barriers readers face when they approach medical research papers. These findings support and deepen prior work on barriers in medical information~\cite{Sommerhalder2009InternetIA, Day2020OpenTT, Nunn2014LaySO}  by illustrating the barriers healthcare consumers face in medical papers, such as uncertainty about where to find relevant information in a paper and an overabundance of jargon (\S\ref{sec:barriers}).
    \item \systemName{}, an interactive reading interface for research papers that integrates existing affordances like term definition tooltips with novel affordances like in-situ plain language summaries of paper sections and a collection of key questions that guide readers to answering passages in the paper with paired plain language answers (\S\ref{sec:design}). 
    \item Evidence from our usability study that these new affordances helped readers quickly find places in a paper that were informative to them. Participants using \systemName{}'s key question index and answer gists had a significantly easier time reading research papers and were more confident they got all relevant information from the papers while retaining a similar level of paper comprehension compared to the typical PDF reader baseline (\S\ref{sec:usabilityResults}). 
\end{enumerate}

\section{Background and related work}

\subsection{Healthcare consumers reading medical research}







Research on consumer health information seeking suggests that trustworthy online health information can empower healthcare consumers, improve clinician-patient interactions, and increase adherence to medical recommendations~\cite{Tan2017InternetHI, Cocco2018DrGI, Broom2005VirtuallyHT, Johansson2021OnlineCA}.~\citet{Tan2017InternetHI} reviewed consumer health seeking behavior and perceptions on using internet information in consultation with clinicians; they found that people did not feel like internet information adversely affected consultations, and that it helped them feel more confident in the consultations and in following clinicians’ suggestions. \citet{Cartright2011IntentionsAA} distinguished two types of health information searching behaviors: evidence-based, which focused on details of symptoms, and hypothesis-based, which focused on understanding a particular diagnosis. In a related setting, ~\citet{Cocco2018DrGI} studied how people search for health information while in an emergency room, showing that many searched for information online on trusted sites like university or hospital websites. ~\citet{Kivits2006InformedPA} explored why healthcare consumers search the internet for medical information, finding that the motivations for searching included helping oneself and filling in missing information from their clinician.~\citet{Choudhury2014SeekingAS} studied health searching and sharing behavior on search engines and social media, finding that search engines are often used for serious medical conditions, but social media can be used to share information about more benign symptoms or conditions. Work has also studied how medically concerned users search for health information online~\cite{Philipp2014InteractionsBH} and how online searching can lead to real-world healthcare utilization~\cite{White2014FromHS}.

While the internet is a good source of consumer health information, there are also many barriers to interacting with this information~\cite{Sommerhalder2009InternetIA, Storino2016AssessingTA}. \citet{White2014ContentBI} analyzed top search results for common health information queries and found that top search results returned for health interventions skewed positively, meaning that more search results said that an intervention will help a condition than suggested by medical evidence.~\citet{Sommerhalder2009InternetIA} found that healthcare consumers searching for information online also struggled with information overload. Information overload can be caused by searches returning unrelated results (e.g., searching a particular symptom and getting results about different diagnoses or home remedies), complex text, or different trusted sites providing contradictory guidance~\cite{Sommerhalder2009InternetIA, Storino2016AssessingTA, Kalavar2021EvaluationOA, Baskin2020GapsIO}. Most people could not resolve these issues themselves, instead needing to discuss the information during consultations with their clinicians~\cite{Sommerhalder2009InternetIA}.  

While many people start out on consumer-facing sites, medical literature is an important source of highly specific, up-to-date information for them~\cite{Zuccala2010OpenAA}. In 2005, the NIH established an open access policy in part to encourage ``individuals [to] become educated consumers about their healthcare and related research, and to consult with healthcare professionals for specific guidance.''~\cite{nationalpolicy} Subsequent research has shown the public benefit of this open access policy, such as improved access to new research findings for healthcare workers and consumers~\cite{Tennant2016TheAE}. While the traditional debate for open-access journals have focused on wider dissemination within research communities, there is an increasing recognition that public stakeholders, including advocacy groups and healthcare consumers, can effectively make use of primary medical research findings~\cite{Day2020OpenTT, Epstein1997ImpureSA}. Indeed, there is a movement in the medical community to involve patients more in the research process, including understanding lab reports~\cite{NAP25094}, reviewing research papers~\cite{Richardsg3726} and leading research efforts~\cite{McCorkell2021PatientLedRC, Nair2012InteractiveJA}. 

At the same time, medical research, and scientific research more broadly, present unique barriers to readers without research expertise \cite{Mnchow2020WhatDI}. \citet{Britt2014ScientificLT} argued that science literacy is the ability to evaluate scientific texts effectively, but that this is challenging due to complex arguments and unfamiliar text structures. \citet{Bromme2014ThePB} highlighted hurdles that the general public face when reading scientific information, including the ability to determine what is relevant and lack of domain expertise.~\citet{Day2020OpenTT} outlined additional barriers specific to searching through medical research, such as lack of adequate scientific literacy, the potential to draw inaccurate conclusions from the findings, and fraudulent journals without sufficient peer review. ~\citet{Nunn2014LaySO} interviewed healthcare consumers on reasons for accessing medical literature and their response to lay summaries written for medical papers. They found that readers appreciated the lay summaries, but often wanted to read the article themselves anyway. At the same time, other work has found that lay summaries help improve reader comprehension compared to journal abstracts~\cite{Kerwer2021StraightFT}. Our project illustrates how interactive reading interfaces can make medical research papers accessible to healthcare consumers through a novel interactive system, \systemName{}. 




\subsection{Interactive reading interfaces}

\systemName{} draws inspiration from prior affordances in interactive reading systems that have used term definitions~\cite{Head2021AugmentingSP}, question answering~\cite{Zhao2020TalkTP, Chaudhri2013InquireBA}, and guided reading~\cite{Dzara2019MedicalEJ} to support reading medical text~\cite{Marshall2016RobotReviewerEO, Brusilovsky1998AdaptiveNS}, dialogue~\cite{li2021hierarchical}, news~\cite{Bohn2021HoneAY}, and search results~\cite{CollinsThompson2011PersonalizingWS}. Inquire Biology \cite{Chaudhri2013InquireBA} is a biology textbook augmented with artificial intelligence (AI) features to support student learning. The textbook allows students to view concept definitions and ask open-ended questions about information in the textbook. If students are unsure of what questions to ask, the textbook also recommends possible questions based on highlighted passages.  In another resource for students, \citet{Dzara2019MedicalEJ} introduced a new method for conducting reading groups that required no prior reading preparation through developing questions about a paper's methodology and findings. They found that these interactive discussions can help pediatric residents analyze medical papers effectively. Also in the clinical context, UpToDate \cite{uptodate} provides expert-written summaries of current research for healthcare providers. 

In the context of reading research papers, \citet{Head2021AugmentingSP} introduced ScholarPhi, a PDF reader that surfaces position-aware definitions for terms defined in a paper (Nonce words) and features for revealing these terms across a paper. In a usability study, researchers were able to read papers more easily using the interface. \citet{Zhao2020TalkTP} introduced ``Talk to Papers,'' a natural language question answering system for exploring research papers. ``Talk to Papers'' allows users to query papers with natural language questions and provides passages where answers are taken from. Other work has explored tools for adaptive summarization in news articles~\cite{Bohn2021HoneAY}, evaluating research literature \cite{Letchford2017AFE, Marshall2016RobotReviewerEO}, navigating concepts within a paper \cite{Abekawa2016SideNoterSP, Jain2018ContentDE} and providing reading guidance in textbooks~\cite{Weber2015ELMARTA, Brusilovsky1998AdaptiveNS}. There are also interactive systems for collaborative reading of research papers, such as Fermat's Library \cite{fermatslibrary}, which provides community annotations on popular research papers, and Hypothes.is \cite{hypothesis}, which allows users to annotate and share annotations on any webpage. 

In contrast to previous reading interfaces for research papers that focus on clinicians, researchers, or students, this project focuses on interactions to make papers understandable to healthcare consumers. There are key ways in which previous designs would not support these envisioned readers. Medical research text is so jargoned that a reader has to invest considerable effort learning the background knowledge to understand it. Previous interfaces that assume readers know what important questions to ask~\cite{Zhao2020TalkTP}, where to look for their answers~\cite{Chaudhri2013InquireBA} or know how to make sense of definitions of terms within a paper~\cite{Head2021AugmentingSP, Jain2018ContentDE} can make reading exceedingly difficult for these readers. \systemName{} goes beyond the typical capabilities of interactive readers to instead help readers understand where to find information of interest in a paper according to the language they already know. To do this, the system incorporates plain language alongside original paper content.   


\subsection{AI for scientific text processing} 

\systemName{} leverages recent gains in natural language processing (NLP) for making medical information more understandable to the public, specifically healthcare consumers~\cite{DemnerFushman2016AspiringTU, Wang2021PretrainedLM}. The research most salient to \systemName{} are automated term definition or replacement~\cite{Veyseh2021MadDogAW}, plain language summarization~\cite{Devaraj2021ParagraphlevelSO}, and consumer biomedical question answering~\cite{Abacha2019AQA}. In addition, we discuss here writing tools to encourage plain language~\cite{Gero2021SparksIF}, as the underlying techniques for powering such systems are similar to those leveraged by \systemName{} (e.g., generating plain language). \systemName{} integrates these advancements in its implementation to show the promise of such methods in supporting healthcare consumers in a user-facing interface and indicate the potential of scaling this reading experience across the scientific literature. 

\citet{Veyseh2021MadDogAW} presented a web-based system for acronym identification that works in the biomedical, scientific, and general domain and~\citet{murthy-etal-2021-personalized} explored how to define scientific terminology with terms recognizable to a specified reader.~\citet{Devaraj2021ParagraphlevelSO} introduced a new dataset of healthcare consumer summaries for clinical topics and a trained model for simplifying medical text.~\citet{Guo2021AutomatedLL} used plain language summaries to train a model for generating summaries of biomedical text. ~\citet{Abacha2019AQA} collected a dataset of consumer health questions from NIH websites and developed methods for automated answering of these questions. ~\citet{Mrini2021AGS} introduced methods to improve answer recall for long and complex consumer medical questions. ~\citet{Gero2021SparksIF} used generation models to help researchers author ``Tweetorials,'' a threaded tweet meant to inform a general audience about a scientific concept on Twitter~\cite{Breu2020FromTT}. Other work has introduced writing tools to help journalists~\cite{Kim2015DeScipherA} or clinicians write using simpler terms~\cite{Van2020AutoMeTSTA, Leroy2013UserEO, Qenam2017TextSU}, simplify text by replacing jargon with more common terms~\cite{Bingel2018LexiAT, Paetzold2016AnitaAI, laban2001kis}, simplify e-prescription and medical instructions~\cite{Li2020PharmMTAN, Cao2020ExpertiseST}, and automatically classify the questions that healthcare consumers ask \cite{Roberts2014AutomaticallyCQ}. \systemName{} draws on this active research to improve access to medical papers. \S\ref{sec:implementation} discusses in depth the adaptations needed to make this research provide useful output for healthcare consumers reading medical research papers.

\section{Observations of non-expert readers}
\label{sec:barriers}

Prior work on reader barriers have focused on consumer health information~\cite{Sommerhalder2009InternetIA}, scientific research in other domains~\cite{Mnchow2020WhatDI}, for students~\cite{Shanahan2011AnalysisOE}, or searching through medical literature~\cite{Day2020OpenTT}, but it is unclear how these barriers manifest for non-experts reading medical research papers. To gather more direct and comprehensive evidence of barriers for this population, we conducted a think-aloud reading study.

\subsubsection{Participants \& recruiting}\label{sec:barriersRecruitment}


We wanted to observe the barriers faced by healthcare consumers when reading medical research. However, the timing of these reading episodes was hard to predict, making it difficult to observe authentic reading experiences. As a compromise, we developed scenarios based on interviews with four healthcare consumers who had prior experience reading medical research and two healthcare providers who had discussed findings from medical papers with their patients. Healthcare consumers and providers were recruited through our personal and professional networks and by referral from other interviewees. More details on these interviews are in Appendix~\ref{app:motivations}. We then recruited participants without medical or research expertise to walk through these scenarios. We provided these participants with a primer about a medical condition and allowed them considerable agency in how they approached the reading task.




We recruited participants who had no experience in the medical profession and in undertaking research via Upwork, a crowd-work site for hiring freelancers. We listed our job under both ``Editing \& Proofreading'' and ``Customer Research'' (i.e., workers partaking in user surveys) to attract a broad sample of workers with varied degrees of reading and writing experience. All participants were paid US\$15 for the hour-long study.\footnote{This is above the federal minimum wage of US\$7.25 and above the region's minimum wage.} We discuss possible limitations to this recruiting strategy and the presence of a paid timed task in \S\ref{sec:limitations}.


A total of 12 participants completed the study (T1--12). Of these participants, 11 had completed college and 5 had completed professional or graduate school. 11 participants had taken 3 or fewer STEM courses since high school.


\subsubsection{Procedure}\label{sec:barriersProcedure}

In the study, participants were given a scenario about a fictional diagnosis representative of common but serious medical conditions (e.g., a herniated disc) with a goal for reading medical papers (e.g., finding new treatments). To ensure participants were equipped with some prior knowledge before approaching papers, they first read a consumer health webpage (MedlinePlus) about the medical condition. This MedlinePlus step was meant to more closely approximate realistic circumstances, in which a participant would have received some information from their doctor about their diagnosis.



We designed the scenarios such that participants would benefit from the additional information found in research papers. To uncover a comprehensive set of barriers, we created four scenarios varied across the following dimensions: diagnosis, demographics (i.e., common or uncommon for a diagnosis), relationship to patient (i.e., patient vs. caretaker), and motivation.

There were two possible diagnoses for each scenario: a herniated disc or systemic lupus erythematosus (SLE, also called Lupus). These diagnoses were selected because they are relatively common and represent serious, long-term issues for a patient. Motivations were: learning background-specific information, becoming aware of emerging treatment options, and comparing treatment options. These scenarios were validated as realistic by a healthcare researcher familiar with consumer health. For more information on these motivations, see Appendix~\ref{app:motivations}.


Participants were randomly assigned into one of the four scenarios. Each scenario was assigned to the same number of participants. After reading a description of the scenario, participants read the MedlinePlus page on the diagnosis then browsed a list of 11 research articles related to the diagnosis. To make these papers representative of the sort healthcare consumers would find in their own searches, we selected only from PubMed articles linked from the MedlinePlus page. We selected papers that were 1) review articles or randomized control trials and 2) relevant to the scenarios (e.g., covering possible new treatments). Papers varied in how relevant they were for a scenario (e.g., some papers covered treatments not clinically available), though all papers had some relevance to the scenarios. While in real-world health information seeking scenarios, readers would undoubtedly come across irrelevant information~\cite{Sommerhalder2009InternetIA}, the study's focus was on barriers in reading papers rather than searching through papers and determining their relevance. Participants chose which papers to consult, which permitted us to see how the contents of a paper affected a participant's choice to read it deeply. Most participants had enough time to read one or two papers (all were asked to read at least one).  






Participants were provided a total of 40 minutes of reading time, split between the MedlinePlus summary page and the papers they chose to read. Participants thought aloud or wrote down any barriers they had while reading. They were prompted for this information every 5 minutes if they had not already volunteered it. The researcher present would sometimes ask participants to elaborate on these barriers. Following the reading, the researcher interviewed participants on what was difficult about reading the research articles and what tools they wish they had to help. After the interview, participants filled out a questionnaire to report their medical literacy and prior research experience.

One author conducted a thematic analysis of the think-aloud and questionnaire data to identify barriers to reading. In multiple meetings, the one author discussed the themes and described evidence with the other authors, refining these themes with input from the other authors. In addition, the other authors confirmed themes by observing recordings of the sessions. We grouped these themes into a set of core challenges, that, if resolved, would help readers make better sense of medical research papers.


\subsubsection{Findings}\label{sec:barriersFindings}



\begin{table*}[]
\centering 
\begin{tabular}
{L{20mm}L{55mm}L{36mm}L{26mm}}
\toprule
 \textbf{Barrier} & \textbf{Description} & \textbf{Example}  & \textbf{Readers}     \\ \midrule  
 \\[-5pt]
 Unfamiliar terminology  & 
 Encountering technical jargon from medicine and biomedical research  & \textit{``What does this word mean?''}
 & T1--3, 5--8, 10--12 
 
 \\ \\
 
 Dense text &  Facing sections of dense text arising from a plethora of jargon & \textit{``I am not going  to act like I understand what any of this means.''} & T1--8, 11--12  \\ \\ 
 
 Knowing what to read & Allocating considerable energy to sections with limited importance  & \textit{``Why did I waste all that time trying to understand what that was?''}  &  T1--3, 5--12\\ \\
 Searching for answers & Asking specific questions but not finding the answers easily  &  \textit{``Where does it talk about the results?''}  & T4, 6, 9--10, 12  \\ \\ 
 Relating findings to personal circumstances & Relating information in the paper to their personal experience  &  \textit{``I would love to know how someone with the same demographics as me responded to this treatment''}  & T2, 5, 8--9, 11  \\
  \\[-5pt]
\bottomrule

\end{tabular}
\caption{Five barriers readers encountered when they sought an understanding of medical research papers without having prior medical research experience. All barriers were caused, or exacerbated by, their lack of expertise in medical research. 
}\label{tab:barriers}
\end{table*}


Our study illustrated barriers readers face when reading medical research papers. The barriers were: unfamiliar terminology; dense text; knowing what to read; searching for answers; and relating findings to personal circumstances. Table~\ref{tab:barriers} lists these barriers. Below we illustrate how these barriers manifested for non-experts reading medical papers, confirming the presence of these difficulties and highlighting concrete instances of difficulties that inform opportunities for design.



\paragraph{Unfamiliar terminology} 

Nearly all (T1--3, 5--8, 10--12) participants mentioned struggling to make sense of the information in the papers because of medical terminology or acronyms that they did not know. These terms ranged from only appearing in some areas of biomedical research (e.g., ``therapeutic peptides'') to commonly used medical terms (e.g., ``comorbidities,'' ``meta-analysis''). The two participants that did not mention struggling with specific medical jargon (T4 \& 9) often skimmed over these terms or were able to infer them from context. Interestingly, while others reported medical terminology as a barrier, they still made some sense of an article without knowing terms by making assumptions about the terms' meanings. At the same time, some terms had meanings that were integral to understanding an article. Incorrect assumptions about these terms could mean misunderstanding the article (T6 \& 10). For example, T10 did not know that ``in vitro'' referred to pre-clinical, non-human studies. They only realized this after reading the majority of the article, which dramatically changed their perception of it's usefulness (i.e., that none of the studied drugs were in clinical trials). 

While terminology is a common barrier in scholarly communication~\cite{Martnez2021SpecializedTR}, past interactions to address it present additional issues for our reading context. Past work has addressed researchers not knowing terms in a paper by providing definitions of terms based on earlier references in a paper~\cite{Head2021AugmentingSP}. There are two issues with such an approach for our reading context: (1) the sheer number of terms could make it difficult for a reader to know which are important 
and (2) there is no guarantee a reader in our context would understand references drawn from the paper, considering that almost all text in medical papers has technical jargon. These issues suggest that a different approach to defining terminology for our envisioned readers is needed.



\paragraph{Dense text} 

While participants could ignore individual terms, such as T4 \& 9, sentences were so filled with these terms, and paragraphs were so filled with these sentences, that participants were overwhelmed by passages of dense text (T1--8, 11--12). This dense text included unfamiliar terminology, but also statistics and complex wording or arguments. Because of the amount of text in the articles and the high cost of reading any of it, participants were quickly overwhelmed. As T8 put it, ``Honestly reading that stuff it was\ldots{}overwhelming just how much terminology I didn't know to start off with\ldots{}It's not like I didn't understand it at all, it was just hard to follow because I had to keep going back, like `Oh what does that acronym mean?' '' T8 was reading a section containing multiple acronyms defined earlier in the paper, including `QoL', `DORIS remission,' and `SLEDAI.' The beginning of one paragraph reads as such: ``In some cases, modifiable causes like anaemia or hypothyroidism may be found, but in most patients, fatigue is unexplained\ldots{}In contrast, SLEDAI or BILAG do not correlate with fatigue.''~\cite{Kernder2020ThePP} T5 expressed a similar sentiment when describing a results passage they were reading: ``I am not going  to act like I understand what any of this means\ldots{}I would have to take the time to understand what these terms mean.'' Continuously having to reference earlier sections of a paper, or searching for term definitions on the internet, can be a major distraction, especially when multiple terms appear in a single sentence. Multiplying this by every sentence in a medical paper creates a categorically different barrier than one term might present. 

Dense text is a barrier that every reader has encountered when learning to read in a new language or domain and is a core motivation for text simplification research. The nuance to this barrier in the context of medical research papers is that readers might have little interest or capacity mastering the language and norms of a particular paper, given that other papers they might read could use different language, and that they may be pressed for time and emotionally and mentally drained from handling their diagnosis.






\paragraph{Knowing what to read} 

Of the 12 participants, 11 (T1--3, 5--12) had a difficult time knowing if a paper held relevant information and invested intense reading effort to determine this. They read papers exhaustively top-to-bottom, reading most of the text, spending time making sense of dense results sections and descriptions of statistical analyses that later they had no use in understanding (T2--3, 5--8). Much of the dense text participants reported struggling with (discussed in the previous barrier) ended up being in sections that they later discovered were less important to read (e.g., a detailed statistical results section). 

One clear example of this was T5, who reported struggling to read the entire first paper they selected because they wanted to do their due diligence by understanding the results. After getting to the discussion they realized that it provided an accessible overview of the results, so for future papers they ignored the technical results sections. As they explained, ``the results, which in my mind would be the first place I would want to go to\ldots{}are very technical and I am not going to know what that means\ldots{}so a general discussion of the results will be more helpful\ldots{}knowing what I know now I would probably skip the results section.'' This quote highlights that non-expert readers lack the knowledge of what they should--and shouldn't--read in a paper, leading them to take much longer learning what a paper has to offer. Other participants had similar experiences as T5, though did not quickly determine what the best passages were for them to read (T2--3, 6--8). 


While some used a paper’s introduction to determine how useful a paper would be, many participants did not trust their ability to know what a paper would contain without exhaustively reading it (T3, 6--8). T6 and 8, for example, both suspected that certain papers would not be useful after reading the abstract or introduction, but continued reading the papers because they hoped they would still find something that was helpful. As we will discuss more in the next barrier---searching for answers---sometimes there was indeed information not surfaced in the introduction or abstract that participants wanted to know, such as low-level details on participant demographics. Participants could invest immense effort to determine if a paper contained this information. In the case of T6, they spent 40 minutes reading a single paper. In another case, T7 reported that they suspected there was useful information in a paper, but it would take them too much time to find it. T3 provided a similar sentiment of wanting a way to know exactly what to read first in a paper: ``I would love some sort of\ldots{}thousand foot-view, which is kind of what I needed in the beginning. Make [the paper] less designed for doctors, and make it more patient friendly, where you are less overwhelmed by all the information all at once, where you can search it out in smaller bites.'' When asked to elaborate, T3 explained that the smaller bites of information could provide high-level findings that they could follow-up on for more details if they were interested. It is worth noting that some biomedical papers do structure abstracts with high level summaries of all sections first or include article highlights at the beginning of the paper, which could help non-expert readers as well as scientists reading these papers. 






\paragraph{Searching for answers}

Participants in our study had specific information they tried to find in the paper, but struggled to do so (T2, 4, 6, 9--10, 12). In contrast to the previous barrier where participants struggled to know what to read in a paper, sometimes participants knew what they wanted to read, but couldn't find this in the paper. The two most salient examples of this barrier were searching for patient demographics and previous treatment options. T2 tried to find information on specific demographic groups in the study to see if they matched their scenario. They had to read through the entire article to find a table with patient demographics and a single sentence within the discussion section making reference to the patient group most relevant to them. Abstracts also did not talk about study demographics or current best practices for treating an illness. Introductions would often include useful information, but it was hidden in background paragraphs or quickly mentioned before moving on to the novel results. Participants therefore had to sift through headers and paper sections, making sense of unfamiliar terms and dense text (two previously discussed barriers) while trying to determine if each sentence was relevant to them. 



\paragraph{Relating findings to personal circumstances} Some participants also wanted additional information from the papers that were personally relevant to them (T2, 5, 8--9, 11). T2 and 8 imagined a tool that could explain how a treatment would affect them, such as by providing patient testimonials for treatments in the paper or results for slices of patients based on demographics. For example, T2 read a paper that reported a 60\% reduction in pain after a surgery, but they wanted to know whether patients regretted the surgery or would recommend it. They also wanted results for a slice of patients most similar to their hypothetical scenario, a 20 year-old male smoker, but the paper only presented average reductions across all patients. T5 found it helpful when an article made reference to the monetary cost of different treatments as a way of referencing patient experiences, though this only happened in one paper. While this personally relevant information was not the goal of the research papers, participants sought this information as a way of relating the information in the paper to their own lives. 
\newline{}
\newline{}
These barriers are unique, or uniquely challenging, to our envisioned readers, necessitating a novel approach to ameliorating them. Past interactive reading systems for research papers have assumed readers have extensive domain knowledge, are able to make sense of paper text as a way of resolving unfamiliar terms~\cite{Head2021AugmentingSP}, know what are the right questions to ask of a paper~\cite{Zhao2020TalkTP}, and understand the basic structure of a paper~\cite{Abekawa2016SideNoterSP}. In contrast, the barriers we identify illustrate that these assumptions do not hold for non-expert readers. Below we discuss how our system, \systemName{}, seeks to address these barriers using plain language (gists) and a collection of key questions as a reading guide, both novel techniques in the context of interactive reading systems for research papers.

\section{\systemName{}: Reading support for medical research papers}
\label{sec:design}



\systemName{} makes medical papers approachable to non-experts. Unlike other systems in the augmented reading space for research papers, \systemName{} focuses on the barriers of non-experts, such as knowing where to invest reading effort. To address this reading context, \systemName{} integrates existing features like term definitions with novel navigational guidance and reading support through a Key Question Index and Answer Gists. 

We focus on four of the five barriers discussed in \S\ref{sec:barriers}: unfamiliar terminology, dense text, knowing what to read, and searching for answers, because these were the most common barriers we observed that hampered readers' ability to get useful information from the papers.  In contrast, relating findings to personal circumstances reflected a desire for additional information and was less focused on understanding information in the paper itself.

We followed an iterative design process for developing \systemName{}. A total of 8 participants used 2 early prototypes of \systemName{} in qualitative usability evaluations. Participants were recruited from our institution, our professional networks, and Upwork. These evaluations lasted one hour each. The iterative design is described in more detail in Appendix~\ref{app:iterative_design}.

One finding from the iterative design we would like to highlight here is the need to supplement, rather than replace, original paper text. We observed participants double checking generated plain language (the gists) with the original text. When asked their reasons for doing so, participants mentioned generated text being vague or wanting to confirm information with the original paper. NLP systems are imperfect (e.g., by generating inconsistent information~\cite{Maynez2020OnFA}) and these observations highlighted the risk of relying solely on generated content. Because of this, in \systemName{}'s design all gists were placed as close to the original text as possible without overlapping, and gist content was provided on-demand, rather than initially displayed along with the paper, to encourage readers to focus on the paper and only pull from the gists for supplemental information. We discuss future designs to further encourage reading original paper text in \S\ref{sec:designImplications}.


\subsection{\systemName{} design}


Based on feedback from the iterative design process, \systemName{} was designed with the following features: 




\begin{enumerate}
     \item \textbf{\defs} -- Tooltips provide definitions of unfamiliar terminology from the open web.
    \item \textbf{\plainSum} -- In-situ plain language section summaries support readers' understanding of dense paper text.
    \item \textbf{Key Question Index} -- Key questions in the sidebar guide readers
    to relevant answering passages.
    \item \textbf{Answer Gists} -- Plain language summaries of the answering passages help readers understand the important information contained in these passages. 
\end{enumerate}


\systemName{} supports healthcare consumers in making sense of medical research papers. To illustrate how \systemName{} is designed towards this goal, we describe how a fictional reader, \reader, leverages \systemName{} to achieve their goal of finding more information about new treatment options. \reader is a first-time reader of medical literature and therefore might differ from some readers who have become familiar with medical terminology because of prior efforts to understand a chronic condition. That being said, we believe that \systemName{}'s features that highlight useful information in a paper can support first-time as well as regular, non-expert readers.


 
 
\reader was recently diagnosed with Systemic Lupus Erythematosus (SLE, also called Lupus), an autoimmune disease. Currently their symptoms are mild: some joint pain and tiredness, but symptoms can worsen and become debilitating over time. When \reader discusses treatment options with their doctor, \reader doesn't know if there are treatments the doctor does not mention that \reader would be interested in. To be informed on available treatments, \reader finds a research paper about possible new treatment options, titled:


\begin{quotation}
``Therapeutic peptides for the treatment of systemic lupus erythematosus: a place in therapy.’’ \cite{Talotta2020TherapeuticPF}
 \end{quotation}

After reading the title, \reader has many questions -- \emph{What is the paper about?} \emph{What are therapeutic peptides?} \emph{Are they a possible new treatments for SLE?} -- and begins reading.

\subsubsection{\reader feels overwhelmed while using a default PDF reader}

\reader starts at the first paragraph of the introduction and immediately becomes stuck on sentences like:


\begin{quotation}
\textit{SLE is characterised by a multifactorial pathogenesis, in which the combination of a favourable genetics and the intervention of external agents may induce the chronic activation of the innate (neutrophils, macrophages, complement system) and the adaptive (T and B lymphocytes, plasma cells, auto-antibodies) immune system.}
\end{quotation}

Even though \reader is familiar with several technical terms like ``innate immune system'' and ``chronic activation'' from reading medical pamphlets and other patient-friendly SLE literature, \reader does not know the meaning of many unfamiliar terms (e.g. ``multifactorial'', ``neutrophils'', ``complement system''). Unable to gauge how critical these terms are for understanding the introduction, \reader  looks up every term one-by-one on the internet. The context switching makes it hard for \reader to recover their place in the paper each time. After ten minutes, \reader realizes that this first paragraph merely provides background on SLE. They haven't made it to the second half of the introduction. 




\subsubsection{\defs help \reader resolve technical jargon without distracting from reading}

\begin{figure}[h!]
    \centering
    \includegraphics[width=0.5\textwidth]{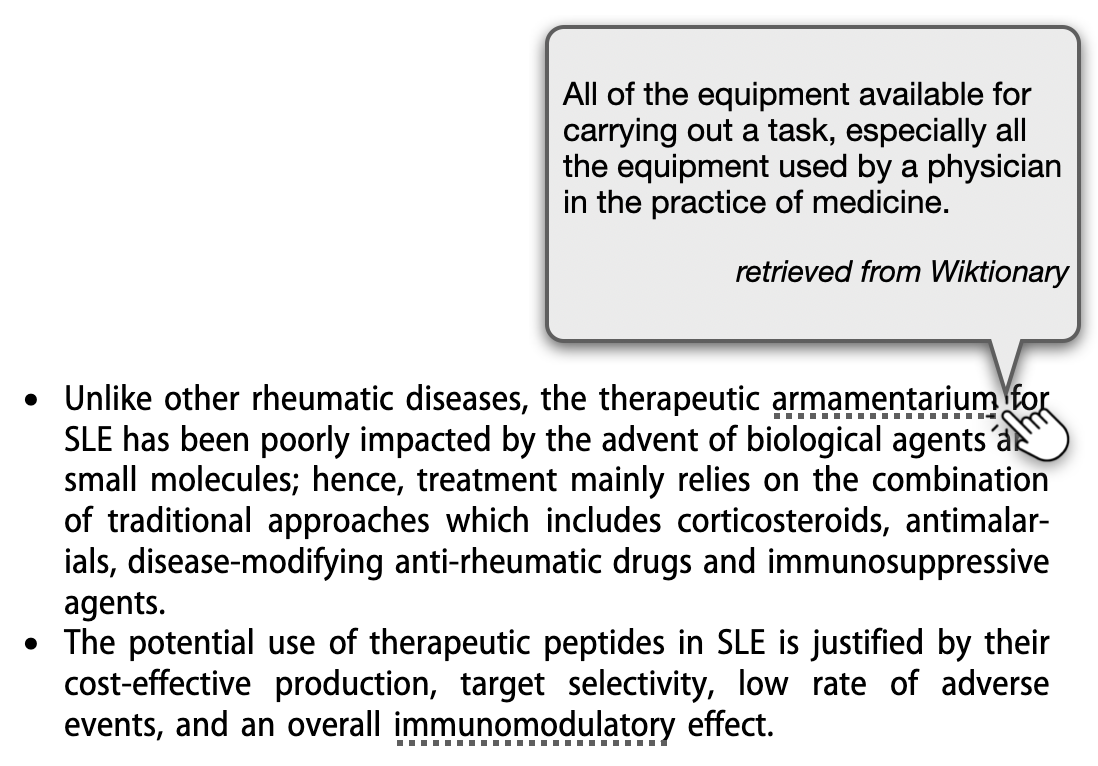}
    \caption{\defs on an example passage with technical jargon. Terms with definitions are underlined (``armamentarium'', ``immunomodulatory''). Clicking a term will open a tooltip with a definition and a reference to the definitional resource.
    }
    \label{fig:TermInteraction}
\end{figure}


\systemName{} provides definitions for unfamiliar terms in the context of the paper so \reader can seamlessly integrate new concepts into their reading. Continuing to read the introduction, \reader reaches another passage full of technical jargon (Figure~\ref{fig:TermInteraction}). In the first bullet, \reader is unsure what ``therapeutic \dotuline{armamentarium} for SLE'' means, preventing them from understanding \emph{what} has been ``poorly impacted.'' Rather than open a new tab to search, they click on the underlined term and a tooltip appears with a short definition retrieved from Wiktionary\footnote{\url{https://en.wiktionary.org/wiki/}} explaining that ``armamentarium'' refers to medical equipment. In the next bullet, \reader sees a list of promising properties of ``therapeutic peptides in SLE,'' but is unsure what ``overall \dotuline{immunomodulatory} effect'' means.  The definition tooltip again helps \reader understand that therapeutic peptides can help control immune functions. \reader continues reading, using the tooltips to resolve unfamiliar terms, eschewing the need to constantly switch tabs.


\subsubsection{The \plainSum help \reader decide whether to invest in reading dense passages}

\begin{figure}[h!]
    \centering
    \includegraphics[width=0.95\textwidth]{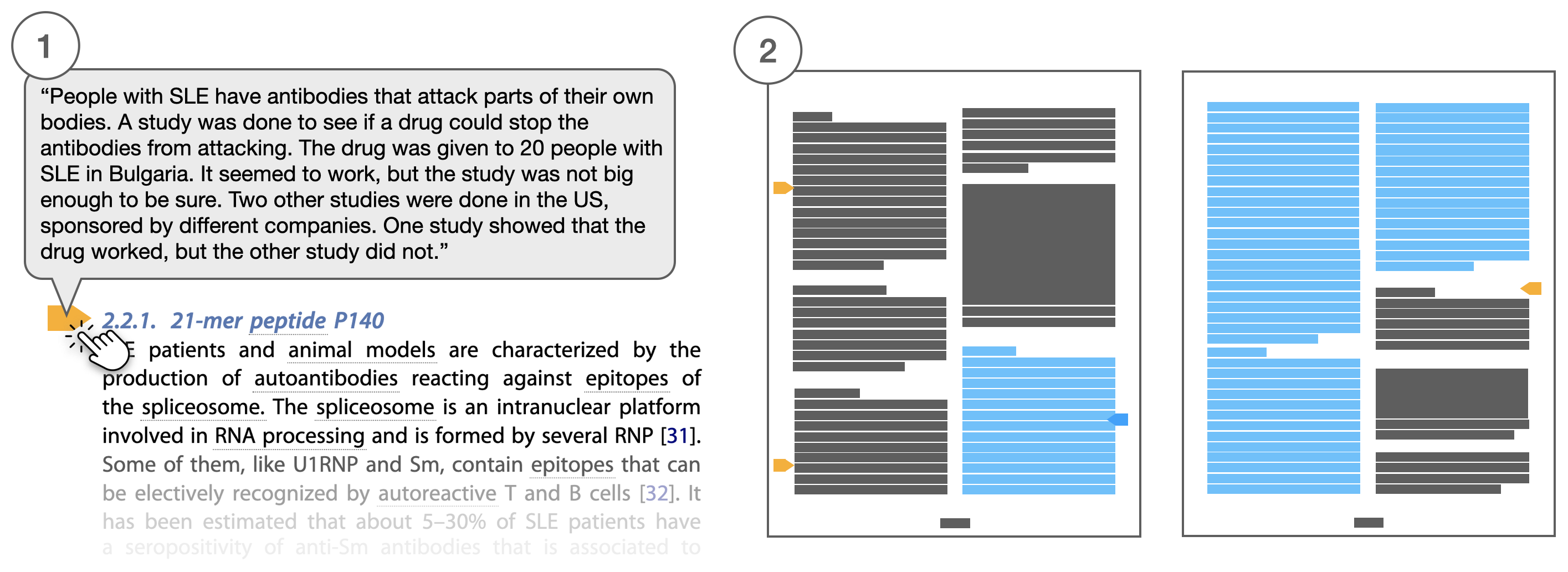}
    \caption{\plainSum on an example passage with dense text. (1) Clicking on a tab indicator next to a section title displays a plain language summary of the section.  (2) Tabs are positioned throughout the paper, providing summaries that can cover a lot of paper content, even across pages.
    }
    \label{fig:SummaryInteraction}
\end{figure}


Equipped with \defs, \reader manages to learn from the introduction that peptides are indeed possible treatments for SLE and wants to learn more.  This particular paper reviews 15 different peptides, each with a dedicated section averaging one page in length; each section includes a description of how the peptide works and its clinical trial results. \reader is motivated to get a high-level sense of each available peptide, but it will require reading 20 pages of dense text. From the introduction \reader gathered that not every peptide has equally promising results and each might be used for different treatments of SLE (e.g., more moderate or more severe cases), so \reader would prefer to only read in depth about the most promising peptides relevant to \reader's mild case of SLE. Skimming through each section, \reader believes some information might be relevant, but it is hard to tell without reading the section in depth. \reader is disheartened that they can't get more details about promising clinical results without going through these walls of text.

\systemName{} makes it easy for \reader to quickly determine what sections are interesting to them and understand the sections with in-situ plain language summaries (\plainSum). \reader clicks on an angled flag next to the section title, and a tooltip appears adjacent to the section text (Figure~\ref{fig:SummaryInteraction}). The tooltip contains a summary of the section stripped of jargon. Rather than sentences like ``SLE patients and animal models are characterized by the production of autoantibodies reacting against epitopes of the spliceosome.'', the summary explains that ``People with SLE have antibodies that attack parts of their own bodies.'' \reader learns from the section gist that this particular peptide has had some good preliminary results, but that further studies have had less successful results. \reader confirms these details by skimming the section and decides this section isn't so relevant to them. \reader uses the \plainSum for the rest of the peptide sections, writing down a few peptides that they are interested in keeping track of, without having to parse all the dense, mostly irrelevant text. \reader completes their reading of these sections in 15 minutes rather than spending hours going through each section in depth. 


\subsubsection{The \keyQs{} help \reader focus on the most important questions and relevant passages.}
\label{sec:keyQs}



\begin{figure}[h!]
    \centering
    \includegraphics[width=1.0\textwidth]{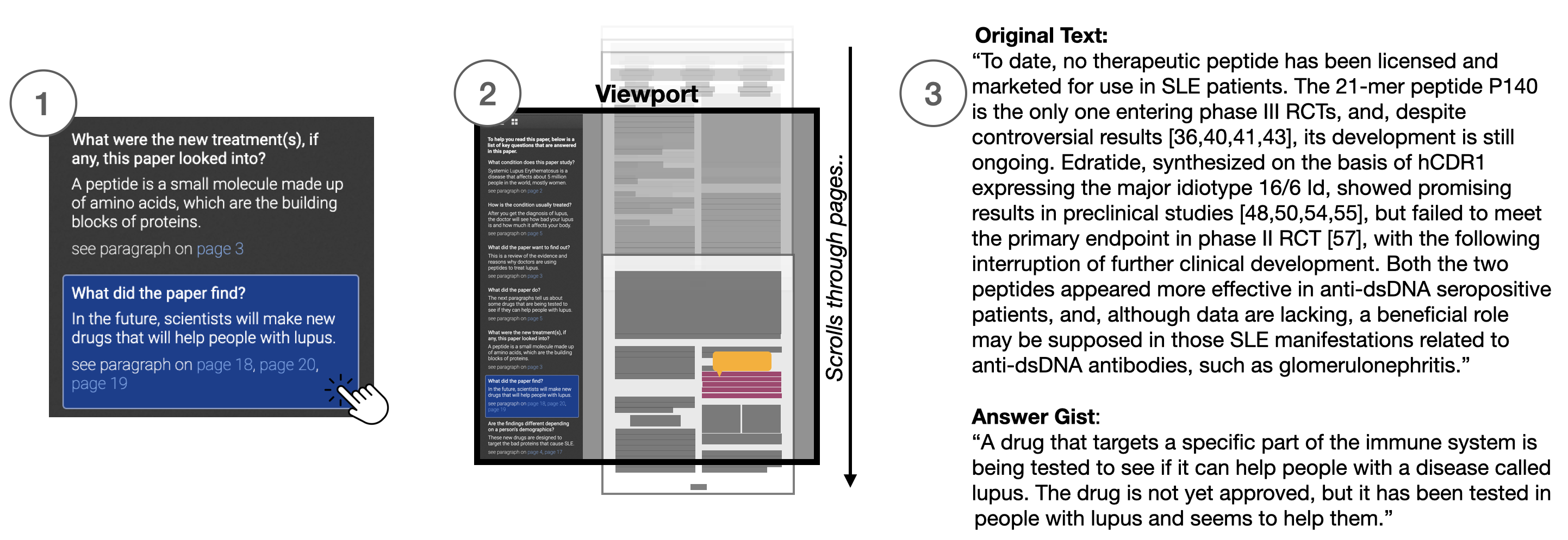}
    \caption{Key Question Index guides readers to answering passages and their Answer Gists. When one of the questions is clicked (1), the interface will scroll (2) to the first answering passage (purple) and display a tooltip (orange) containing the Answer Gist. In (3), we show the simplified Answer Gist alongside the original paper text.
    }
    \label{fig:keyQInteraction}
\end{figure}

\reader gets to the end of the paper using the \plainSum{} to read only some sections in depth, but is worried they might miss important information in the paper because they didn't know to look for it. \reader got a general sense of each section using the \plainSum{} but is curious if there is some information that the general summaries might not have surfaced, especially in larger sections containing lots of relevant information, such as the Discussion or Introduction. 




As an alternative to assessing relevance 
with \plainSum{}, \systemName{} provides \reader with key questions linked to answering passages in the paper along with plain language answers to point \reader to important information. \reader looks to \systemName{}'s sidebar and sees questions about the paper that cover key information, such as ``What did the paper do?'' and ``What did the paper find?'' \reader sees that the question ``What did the paper find?'' hyperlinks to multiple passages within the Discussion (see (1) Figure~\ref{fig:keyQInteraction}). They click on the first link. \systemName{} scrolls through the pages and settles on a highlighted paragraph in the Discussion summarizing the most promising therapeutics peptides (see (2) Figure~\ref{fig:keyQInteraction}). Unfortunately, the answering passage looks dense. As \reader prepares to look up more terms, they notice a tooltip below the answering passage containing a plain language summary (an ``Answer Gist''). This answer gist is a quarter the length of the original paragraph and contains none of the unfamiliar terms (see (3) Figure~\ref{fig:keyQInteraction}). While the answer gist by itself might not contain all the information \reader wants, they can read the original paragraph along with the answer gist, comparing the complex wording with plain language and get a general understanding of the paragraph without being overwhelmed by technical jargon. Similar to the \plainSum{}, \reader can then dive into the original passage with this understanding to get more details. \reader clicks through the rest of the links for the same question, which scrolls them to individual paragraphs in the discussion that cover the most important findings and interpretations of the paper. 

\systemName{}'s key questions also guide readers to the questions they might not know to ask about a paper. Before finishing reading the paper, \reader looks through the rest of the questions in \systemName{}'s sidebar. Each question is accompanied by a one-to-two sentence plain language answer preview and hyperlinks to one or more paragraphs in the paper that answer the question. With only a handful of key questions and short answers, a majority of the questions can be displayed in the sidebar without scrolling so \reader can quickly read all the questions and answers with minimal effort (see (1) in Figure~\ref{fig:keyQInteraction}). \reader sees and clicks on one question they hadn't thought to look for in the paper: ``What are the limitations of the findings?'' \systemName{} scrolls them to a paragraph in the Conclusion saying that not only are therapeutic peptides currently not licensed for clinical use for SLE (which \reader had already read), but also that many of the current clinical trials have mixed efficacy results and that future clinical trials might show more promise with different study designs (which \reader had not already read). \reader is glad they confirmed and deepened their understanding of the paper's limitations with this final question. 

\reader has spent only a few minutes to learn the most important information about the paper for them: these are not treatments they could ask their doctor to prescribe them, but there might be some promising clinical trials \reader could look into. They also feel confident that for future papers they could use this key question sidebar to quickly get a high-level summary of the most important information in a paper.

\section{Implementation}
\label{sec:implementation}
\systemName{} leverages active research in NLP for biomedical question answering~\cite{10.1007/978-3-030-43887-6_64} and plain language summarization~\cite{Guo2021AutomatedLL} to address reader barriers. Below we discuss the implementations powering each feature of \systemName{}. While additional algorithmic advances or human oversight, specifically for ensuring factuality~\cite{Maynez2020OnFA}, are necessary to make deploying such a system safe, our current implementation indicates the potential for \systemName{} to be deployed at scale over the medical literature. 

\begin{figure}[t!]
    \centering
    \includegraphics[width=0.75\textwidth]{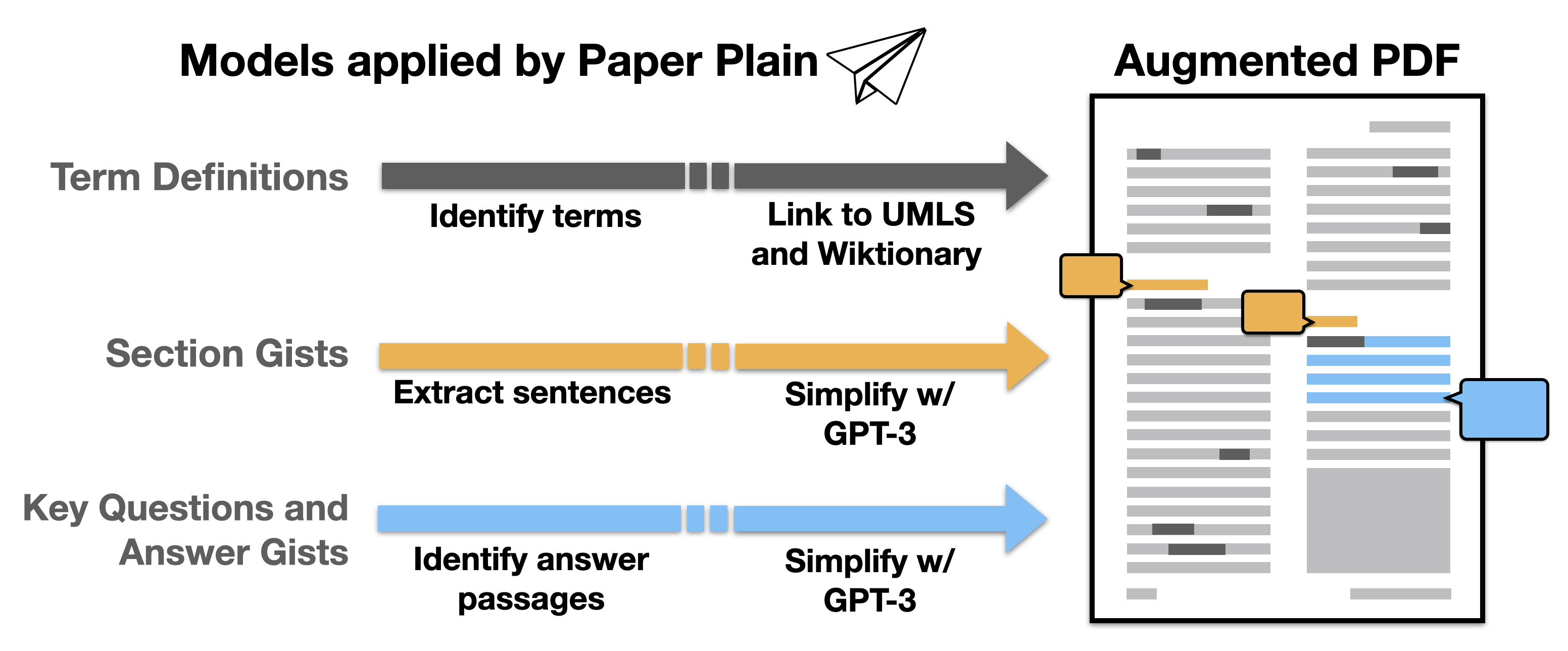}
    \caption{\systemName{} uses machine learning models to add term definitions, section gists, and answer gists to the PDF.}
    \label{fig:system_overview}
\end{figure}

\subsection{\defs{}}

\systemName identifies medical terms in the paper using \texttt{scispaCy} Named Entity Recognition (NER)~\cite{neumann-etal-2019-scispacy} and links these terms to definitions from the Unified Medical Language System (UMLS)\footnote{\url{https://www.nlm.nih.gov/research/umls/index.html}} or Wiktionary.\footnote{\url{https://en.wiktionary.org/wiki/}} The extraction and linking process led to many false positives (e.g., identifying terms like `expert' or `negative'), so we additionally filter terms based on word frequency and length. For both Wiktionary and UMLS, we preserve the bottom 80\% of terms based on word frequency and remove all terms at or above 30 characters (terms over 30 characters were usually ill-formed, for example, containing a citation string or the beginning of the next sentence). We additionally filter all Wiktionary definitions to those containing at least one of the following tags: `medicine', `organism', `pathology', `biochemistry', `autoantigen', `genetics', `cytology', `physics', `chemistry', `organic chemistry', `immunology', `pharmacology', `anatomy', or `neuroanatomy.' 


\subsection{\plainSum{}} We define section gists for the lowest level subsections in the paper (e.g. 2.2.1). To generate section gists, we concatenate the first sentence of every paragraph in a section and generate a plain language summary of it using GPT-3~\citep{Brown2020LanguageMA}. GPT-3 is a pretrained generative model released by OpenAI that has obtained state-of-the-art results on many language tasks using different prompts for generation \citep{Brown2020LanguageMA} and is commonly used for many generative tasks (e.g., generating plain language). We engage in prompt engineering, a common practice for achieving fluent text for large generative models~\cite{Liu2021PretrainPA}, to encourage fluent and specific plain language summaries. We use a GPT-3 prompt adapted from a preset example that OpenAI provides for simplifying text.\footnote{\url{https://beta.openai.com/examples/default-summarize}} We modified the prompt to suggest a fifth grade reading level rather than second grade. We also tested later grades, up to college, but found that the generated text using the fifth grade reading level prompt was the most coherent while still providing some details about the section. Sentences were extracted manually for our prototype system, but could be automatically extracted using PDF parsing methods \citep{Lopez2009GROBIDCA, Shen2021IncorporatingVL}. Using the leading sentence of each paragraph is a common competitive baseline for summarization~\citep{Erkan2004LexRankGL}; we choose this strategy rather than inputting the full section text because GPT-3 is prone to copying the text verbatim when given the full section. 

We observed variations in summary quality, such as hallucinated or incorrect information (e.g., calling peptides a surgery), repeated words or sentences, and copied text from the original passage.  
In these cases, we would regenerate the summaries up to five times and select the most fluent or correct generation. This usually provided a coherent and correct summary, but there were examples of text copied from the original passages that persisted. We discuss generation quality and accuracy further in \S\ref{sec:ethicalImplications}. More details on the GPT-3 prompt are in Appendix~\ref{app:implementation}.



\subsection{\keyQs{}} Key questions were drawn from two sources designed to translate medical findings applicable to patients: the PICO framework~\citep{Richardson1995TheWC} for clinical questions and Cochrane’s guide on writing plain language summaries~\citep{CochranePLS}. Both sources focus on information in medical papers that are relevant to patients and caregivers. We curated 8 questions from the two sources for inclusion in \systemName{}; these are listed in Table~\ref{tab:questions} in the Appendix.


For each question, \systemName{} extracts relevant passages from the paper using an extractive question answering (QA) system trained on BioASQ, a biomedical question answering task~\cite{10.1007/978-3-030-43887-6_64}. Because this QA model extracts single words or phrases rather than full passages, we used the entire paragraph that contains an answer extracted by the model. For our prototype system, we manually labelled sentence boundaries of the extracted answers on the PDF to ensure high quality bounding boxes for display. Recent work has improved the accuracy of automatic sentence bounding box extraction from PDFs~\cite{Shen2021IncorporatingVL}, which could be used to automate this step in the future. We follow prior work on making QA models more robust by including semantically-equivalent variations of questions~\cite{Gan2019ImprovingTR}.

In the system, we highlight the paragraph containing the answer and display an answer gist summarizing the answer. We create answer gists by simplifying the extracted passages using GPT-3~\citep{Brown2020LanguageMA} with the same prompt we use for simplifying section gists. We also include the first 1-2 sentences of the answer gist in the sidebar along with the question.
\section{Usability study}
\label{sec:usabilityStudy}


\systemName is meant to help readers engage with medical research papers important to them. We ran a within-subjects usability study to assess how well \systemName{}'s features meet these goals. 

The study answers the following questions:


\begin{enumerate}
    \item[] \textbf{RQ1}-How did participants use \systemName{}'s features?
    \item[] \begin{itemize}
        \item \emph{Did participants prefer some features over others?}
        \item \emph{Did participants use features throughout the reading session?}
        \item \emph{Did presence of one feature affect usage of another feature?}
        \item \emph{Did participants traverse linearly through a paper or employ a jumping reading strategy?}
        \item[] 
    \end{itemize} 
    \item[] \textbf{RQ2}-How does \systemName{} affect participants' self-reported reading difficulty, understanding, and ability to identify relevant information?
    \item[] \begin{itemize}
        \item \emph{...in comparison with a standard PDF reader?}
        \item \emph{How does providing reading guidance (i.e., the \keyQs features) affect these self-reported metrics?}
        \item \emph{...in comparison with an interface with only non-guidance features (i.e. \plainSum and \defs )?}
        \item[] 
    \end{itemize}
    \item[] \textbf{RQ3} - Do we observe any difference in paper comprehension when participants use \systemName{}? 
    \item[] \begin{itemize}
        \item \emph{...in comparison with a standard PDF reader?}
        \item \emph{What is participant behavior in the presence of incorrect system predictions (e.g., vague information or factual errors in generated gists)}
    \end{itemize}
\end{enumerate}
%

\subsection{Method}

\subsubsection{Participants}

We recruited participants from Upwork using the same recruiting materials as \S\ref{sec:barriersRecruitment}. We again recruited from both the ``Editing \& Proofreading'' job category and ``Customer Research'' to attract a broad sample of workers with varied degrees of reading and writing experience and to remain consistent with \S\ref{sec:barriersRecruitment}. All participants were paid US\$15 for the hour-long study.

A total of 24 Upworkers (9 male, 1 non-binary, and 14 female) participated in the study. Participants' age ranged from 19 to 67 ($\mu=35.04$). All participants had completed college, and a third had completed professional or graduate school. 79\% of participants (19) had taken 3 or fewer STEM course since high school and 92\% (22) had never been involved in publishing a research paper. Similar to \S\ref{sec:barriers}, no participants had professional medical experience.


\subsubsection{Procedure}

The usability study consisted of two parts, each corresponding to a scenario involving a patient with a particular diagnosis---systemic lupus erythematosis (SLE) or a herniated disc---and who was interested in exploring new treatments. The scenarios for each paper were drawn from \S\ref{sec:barriersProcedure}. For each scenario, we selected a single paper (\cite{Talotta2020TherapeuticPF} for SLE and \cite{Bai2021PercutaneousEL} for a herniated disc) for participants to read based on the most common papers readers selected in \S\ref{sec:barriers}.  
Each participant underwent the following study procedure once for each scenario. First, participants read a description of the scenario, the MedlinePlus page about their diagnosis and the associated research paper. Then, they answered questions about the paper. 


Participants read the scenario description and had 2 minutes to read the MedlinePlus page on the diagnosis. They went through a short tutorial on \systemName{} then read the paper for 10 minutes. They were told at 5 minutes and 9 minutes how much time they had remaining. After each paper, participants filled out subjective ratings and multiple choice questions about the paper (covered in \S\ref{sec:measures}). After the two scenarios, participants completed a questionnaire on their demographics, education, and research experience. Following the questionnaire, participants completed a short form on their experience using \systemName{} and what features they found most helpful. A researcher was present for the entire experiment and followed up on these answers with additional probing questions in a final interview. 

\subsubsection{Measures}
\label{sec:measures}

We collected measures for assessing feature usage (\textbf{RQ1}), subjective reading experience (\textbf{RQ2}) and comprehension (\textbf{RQ3}):

\paragraph{Feature usage}

To measure how participants used \systemName{}'s features (\textbf{RQ1}) we collected telemetry data on interactions with \systemName{}'s features (e.g., opening a definition tooltip or clicking on a key question). We report feature usage over the 10 minutes of reading each paper. We determine significant patterns of usage if the majority of participants exhibited this pattern, as observed by researchers present in the experiment and corroborated by the rest of the authors when examining usage data. 


\paragraph{Subjective reading experience}

We collected subjective ratings to understand how \systemName{} affected participants' reading experience. Participants completed the ratings after reading each paper. These included:

\begin{enumerate} 
    \item Reading difficulty: Participants rated their reading difficulty on a 1-5 Likert-style scale based on the question: ``How hard did you have to work to read the paper?'' 
    \item Understanding: Participants rated their understanding of the paper on a 1-5 Likert-style scale based on the question: ``How much do you feel like you understood the paper?'' 
    \item Relevance: Participants rated their confidence they got any relevant information from the paper on a 1-5 Likert-style scale based on the question: ``How confident are you that you got all the relevant information from the paper?''
\end{enumerate}





\paragraph{Comprehension}

While \systemName{}'s primary goal is to support readers in navigating and identifying relevant information in a paper (captured by our subjective reading experience measures), it is also important to ensure that \systemName{}'s affordances do not detract from overall paper comprehension (e.g., by over-simplifying or incorrectly summarizing paper content). We developed multiple choice questions to measure the degree to which participants understood the paper (\textbf{RQ3}). Our goal when developing the multiple choice questions were to ensure that they:

\begin{itemize}
    \item were specific to the individual papers,
    \item were relevant in a clinical context, and
    \item could not be answered directly from the \keyQs{} in \systemName{}.
\end{itemize}

We achieved these goals by writing 15--20 questions for each paper and having two practicing physicians not involved in the study provide feedback on the questions. The clinicians read the papers without \systemName{}, gave feedback on all questions, and selected 5--7 they thought were most meaningful for overall paper understanding and were important in a clinical context. We revised wording on any questions or answers that were unclear or easy to misunderstand according to the clinicians and two additional pilot studies. At the end, we selected 14 multiple choice questions, 7 for each paper. All questions could be answered from text not highlighted by the \keyQs{}. Paper comprehension was measured as the proportion of questions answered correctly. Participants answered these questions after completing the subjective ratings for a paper. 

While it is important to ensure that any augmentation does not negatively impact paper comprehension, it is worth noting that prior work on augmented reading interfaces have not shown significant differences in the number of comprehension questions answered correctly across conditions~\cite{Head2021AugmentingSP, Badam2019ElasticDC}. While observing an improvement in comprehension due to \systemName{} would be an exciting addition to our primary objective of improving reading experience, our goal with this measure was ensuring that \systemName{}'s features did not lead to a loss in comprehension.


\subsubsection{Interface variants}

To understand the impact of \systemName{}'s novel guidance-offering features on readers' experience engaging with medical research papers, we evaluated variants of \systemName{} with and without these features. There were three versions of \systemName{} and one baseline:

\begin{enumerate}
    \item \systemName{} -- The full interface with the \keyQs{}, \plainSum{}, and \defs{}. 
    \item \navGuide{} -- The guidance-focused variant with only the \keyQs{}. 
    \item \inSituExpl{} -- The variant without guidance, providing readers with the \plainSum{} and \defs{}. 
    \item \pdf{} -- A typical PDF reader.
\end{enumerate}


\paragraph{Assignment}

We assigned each participant to two of the possible eight variant--paper configurations. Each participant saw each paper once, and all eight configurations had the same number of assigned participants. All assignments were counter-balanced so that each configuration was experienced as the first or second task the same number of times. 



\subsubsection{Analysis}\label{sec:usabilityAnalysis}

We compared readers' subjective ratings for reading difficulty, understanding, and relevance across the system variants (\systemName{}, \navGuide{}, \inSituExpl{}, \pdf{}) using mixed-effects linear models~\cite{Lindstrom1990NonlinearME} with paper type and system variant as fixed effects and participant as a random effect. Using a mixed-effects model for each measurement, we first conducted $F$-tests for any significant difference across the system variants, and then we conducted $t$-tests for differences in the estimated fixed-effects between all pairs of system variants. More details are in Appendix~\ref{app:stats}.



We conducted a non-inferiority test~\cite{Walker2010UnderstandingEA} to assess \systemName{}'s impact on comprehension. Our goal was to confirm that \systemName{} did not detract from paper comprehension. Prior work has suggested that plain language can overly-simplify scientific findings, risking reader misunderstandings~\cite{Scharrer2017WhenSB, Sumner2014TheAB}. The multiple choice questions were designed to assess general paper understanding, meaning that all questions were answerable from information in the paper that appeared multiple times (e.g., the paper stating its main findings in the abstract, introduction, and discussion). Because \systemName{} was designed to make it easier for readers to access the information in a paper, not to add significant additional information, we did not expect that \systemName{} would dramatically improve comprehension as measured by our questions. However, we wanted to ensure that \systemName{} did not detract from comprehension. The non-inferiority test was conducted using the  \texttt{statsmodels} package in Python~\cite{seabold2010statsmodels} as the lower bound of a t-tost (two independent t-tests).

For qualitative findings, one author conducted a thematic analysis on the observations of the study sessions and discussed themes with the other authors to refine these themes. Themes were identified via open coding and discussed in 3 weekly meetings with all authors. One author coded all interviews, while another author verified the themes in one of the interviews. 





\section{Results}
\label{sec:usabilityResults}
Below we report our findings from the usability study broken down by research question.


\subsection{How did participants use \systemName's features?}
\label{sec:RQ1}

\begin{figure}[t]
    \centering
    \includegraphics[width=1\textwidth]{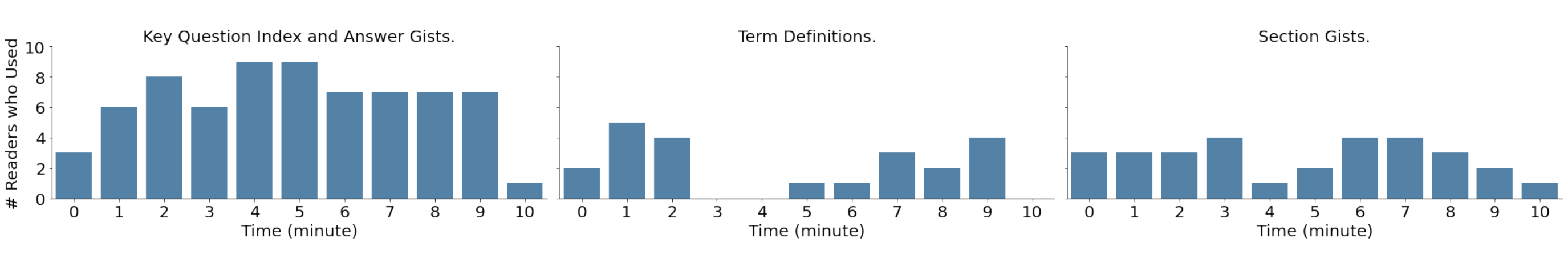}
    \caption{Number of readers who used each feature of \systemName{} for each minute of reading a paper. Plot is for participants with access to all features of \systemName{}.}
    \label{fig:usage_counts}
\end{figure}


\begin{figure}[]
    \centering
    \includegraphics[width=0.8\textwidth]{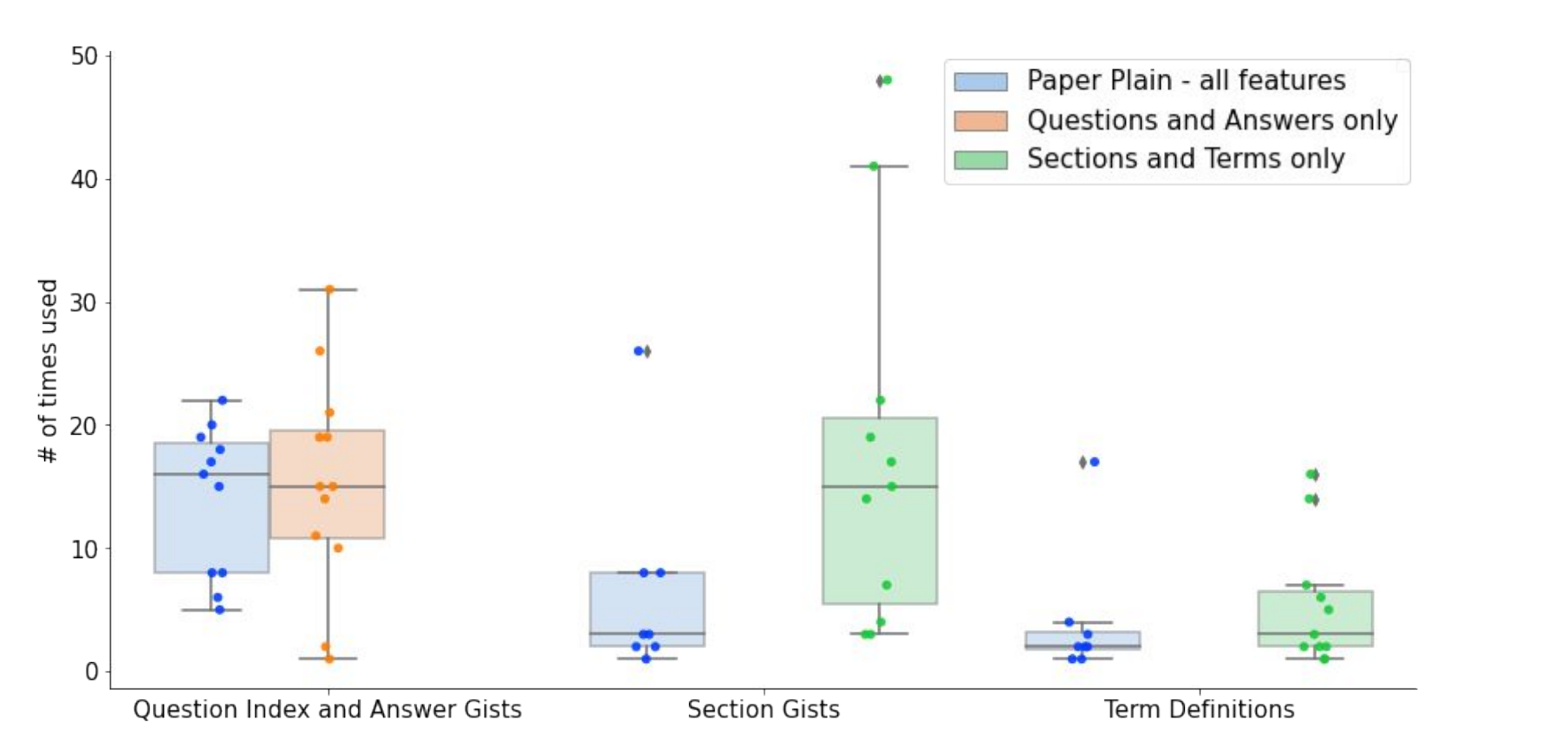}
    \caption{Number of readers who used each feature across the different variants of \systemName{}. Points represent individual readers. For example, all but one reader who had access to all features used Term Definitions more than 5 times, shown by the single blue dot above the rest in the far right of the plot, at the `Term Definitions' tick. Notice the drop in usage of \plainSum{} and \defs{} when all features are available (the blue boxplots). }
    \label{fig:conditionUsages}
\end{figure}

Most participants interacted with all the features of \systemName{} available to them. All participants with access only to the \keyQs{} (\navGuide{}) clicked on at least one Key Question and opened an Answer Gist. Usually they clicked on many more: on average participants with this variant clicked on 15 Key Questions and Answer Gists. 11 out of 12 participants with the \plainSum{} and \defs{} (\inSituExpl{}) clicked on a Section Gist and a Term Definition. On average, participants with this variant clicked on 18 Section Gists and 5 Term Definitions.

When participants had access to all the features they often opted for the \keyQs{}. 11 out of 12 participants with access to all of \systemName{} clicked on a Key Question and opened an Answer Gist, doing so on average 13 times for Key Questions and 14 for Answer Gists. In contrast, only 8 participants with \systemName{} clicked on a Section Gist or Term Definition. Participants that did engage with these latter features also used them much less, clicking on average only 7 Section Gists and 4 Term Definitions. Figure~\ref{fig:conditionUsages} plots the usage of each feature for \systemName{} and illustrates this preference for the \keyQs{} when all features were present. 

Participants used \systemName{}'s features throughout reading a paper. Figure~\ref{fig:usage_counts} plots the number of participants using each feature over the course of reading a paper. There is a slight `warm-up' period for each feature--usually in the first two minutes--where participants used the features less. Usage increases after this initial phase, with sustained interaction that tapers in the last minute of a study. 


Along with this sustained usage we observed changes in reading strategies when participants had \systemName{}'s features compared to when they did not. Most participants with the baseline PDF reader read papers linearly and, similar to what we observed in \S\ref{sec:barriers}, got stuck in dense sections with limited important information (e.g., technical backgrounds) (P2, 5, 6, 10, and 22). For example, P22 did not get to the end of one of the papers because they were focused heavily on understanding the dense methodology and background sections. When told they had a minute left, all but one of these participants (P2, 5, 10 and 22) quickly scrolled to the end of the paper to read the sections there, suggesting that they viewed these sections as important but did not have adequate time to read them.  

All participants with \systemName{} made it to the end of a paper; \systemName{}'s features supported readers in doing so in different ways. The \plainSum{} and \defs{} helped participants understand dense text (P1, 3--5, 7, 15, 18), while the \keyQs{} allowed participants to quickly find text that was informative for them (P2, 4, 7--10, 13, 18--20).

Participants with the \plainSum{} and \defs{} were able to easily make sense of dense passages (P1, 3--5, 7, 15, 18). As P18 explained,  ``It [the \plainSum{}] broke down very complicated medical text into easily understandable terms that helped me to keep up with the article and not skip over the wall of text.'' Participants also used the Section Gists to decide whether or not they wanted to read a section and, when they decided to read the section, as a guide for understanding the complex text (P5, 7). As P7 reported, they ``liked having a brief summary of what to expect so I don't walk in completely clueless.'' This feedback aligns with our design goal for the \plainSum{}, which was to help readers avoid reading an abundance of dense text by giving them a preview of what the text is about (\S\ref{sec:design}). 

In contrast, participants with the \keyQs{} used the questions' guidance to quickly find text that was informative for them by jumping to that information (P2, 4, 7--10, 13, 18--20). A clear example of this was P10, who read through the abstract and introduction of a paper, then opted for using the Key Questions to jump through different sections of the paper. When asked why, they replied that they trusted that the questions would provide them the information they were looking for. 

\begin{figure}[]
    \centering
    \includegraphics[width=0.9\textwidth]{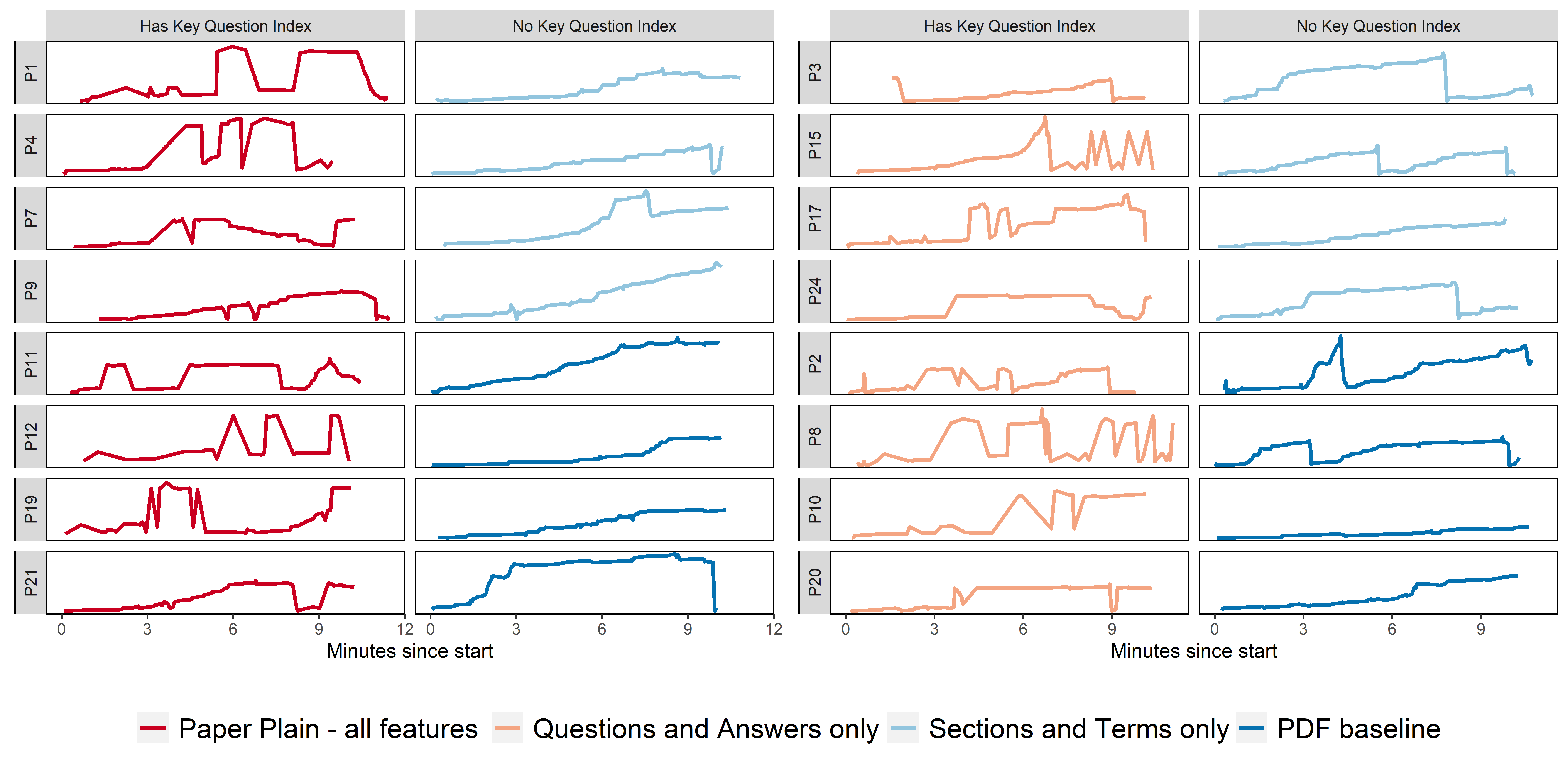}
    \caption{
    Participant's reading behavior differed when reading one paper with the \keyQs{} and another paper without. Each plot's y-axis calculates the vertical position of the participant's viewport relative to total paper length (e.g., the bottom and top of each graph are the beginning and end of the paper, respectively). We observe much more jumping behavior when the \keyQs{} are present.
    }
    \label{fig:exLinePlot}
\end{figure}

How the \keyQs{} encouraged a nonlinear reading strategy is also reflected in where participants spent their time in a paper. Participants with the \keyQs{} jumped back and forth through a paper, shown by the consistent usage of jumping to answers in Figure \ref{fig:usage_counts}, while participants without the \keyQs{} often read papers top to bottom, once through. This behavior is exemplified in Figure~\ref{fig:exLinePlot}, which plots participants' position in a paper over the course of reading with and without the \keyQs{}. 

The jumps afforded by the \keyQs{} also brought participants to places in the paper they spent more time reading compared to participants without this feature who scrolled through the paper linearly. Participants with the \keyQs{} averaged $5.19$ seconds ($\sigma=7.72$) in each paper position before moving, which was $1.85$ seconds longer than what participants spent without the \keyQs{} (average $3.34$ seconds, $\sigma=10.99$). A key motivation for the \keyQs{} was to fast-track readers to the important information in a paper. The jumps in Figure~\ref{fig:exLinePlot}, combined with evidence of more sustained reading, suggest that participants took advantage of this affordance.





\subsection{How does \systemName{} affect participants' self-reported reading difficulty, understanding, and ability to identify relevant information?}




\begin{figure}[t]
    \centering
    \includegraphics[width=1\textwidth]{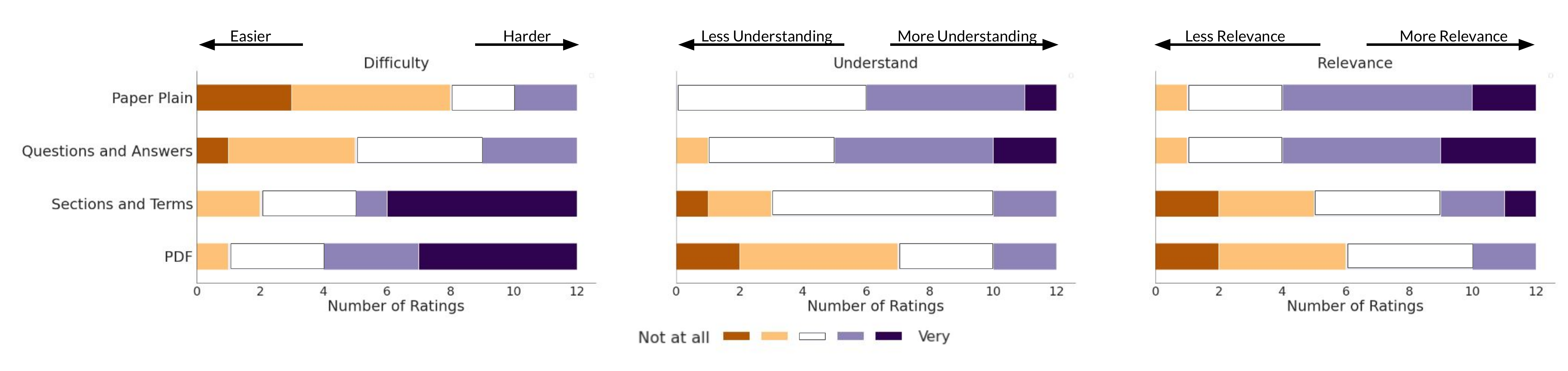}
    \caption{Readers' subjective reading difficulty, confidence in their understanding of the paper and ability to get all relevant information from the paper for different variants of \systemName{}.
    }
    \label{fig:confidence}
\end{figure}

Figure~\ref{fig:confidence} plots the reading difficulty, understanding, and relevance scores for both papers across each system variant, and we observe significant differences between them. This is also reflected in our mixed-effects model $F$-test ($p < 0.001$ for all three measurements after Holm-Bonferroni~\cite{Holm1979ASS} correction). We report estimated fixed-effect coefficients in Appendix~\ref{app:stats} and instead discuss more interpretable results comparing system variants in this section. We report here on medians (denoted $\tilde{x}$) for each subjective rating because ratings were scored on Likert-style scales. 

Table~\ref{tab:pairwise-contrasts} presents the differences in the fixed-effects between all pairs of interface variants. Participants with \systemName{} were significantly more confident that they got all relevant information from the papers ($\tilde{x}=4.00$, $\sigma=0.87$, with $5.00$ being the most confident) and understood the papers ($\tilde{x}=3.50$, $\sigma=0.69$), compared to the PDF reader baseline ($\tilde{x}=2.50$, $\sigma=1.00$ and $\tilde{x}=2.00$, $\sigma=1.00$). Participants with \systemName{} also rated their reading difficulty significantly lower ($\tilde{x}=2.00$, $\sigma=1.06$, with $5.00$ being hardest) compared to participants who had the PDF reader baseline ($\tilde{x}=4.00$, $\sigma=1.04$).


\begin{table*}
\centering
\begin{tabular}{rcccccc}
\toprule &
$PP - QA$ &%
$p$ &%
$PP - SD$ &%
$p$ &%
$PP - PDF$ &%
$p$ \\

\midrule              

Reading Difficulty (1--5)&%
-0.344 &%
0.7481 &%
-1.485 &%
\textbf{0.0011} &%
-1.983 &%
\textbf{$<$.0001} \\

Understand (1--5) &%
-0.104&%
0.9842&%
0.719 &%
0.0866 &%
1.177 &%
\textbf{0.0020} \\

Relevance (1--5)&%
-0.193&%
0.9133&%
0.752 &%
0.0772  &%
1.167 &%
\textbf{0.0030}  \\

\bottomrule

\toprule &
$QA - SD$ &%
$p$ &%
$QA - PDF$ &%
$p$ &%
$SD - PDF$ &%
$p$ \\

\midrule          

Reading Difficulty (1--5)&%
-1.141&%
\textbf{0.0132}&%
-1.639 &%
\textbf{0.0003} &%
-0.498&%
0.4786 \\

Understand (1--5) &%
0.823&%
\textbf{0.0401}&%
1.281 &%
\textbf{0.0008} &%
0.457&%
0.4106 \\

Relevance (1--5)&%
0.946&%
\textbf{0.0183}&%
1.361 &%
\textbf{0.0006} &%
0.415&%
0.5093 \\

\bottomrule
\vspace{0.2ex}

\end{tabular}

\caption{
Post-hoc (two-sided) tests for pairwise differences in fixed-effects estimates between interfaces. \textmd{This table reports the difference in fixed-effects estimates $i - j$ and Holm-Bonferroni-corrected $p$-values~\cite{Holm1979ASS} under our mixed-effects model, where $i$ and $j$ correspond to interface options --- $PP=$ \systemName, $QA=$ \keyQs, $SD =$ Both \plainSum and \defs, and $PDF =$ PDF baseline.  
For example, in the column for $PP - PDF$ and row for ``Reading Difficulty,'' we can
interpret the result as \systemName{} is associated with, on average, 1.983 points lower rating of reading difficulty than a PDF baseline when controlling for participant and paper.  Statistically 
significant $p$-values are bold.  More details about this analysis are in 
Appendix \ref{app:stats}. 
}}
\Description{%
Data table of pairwise comparisons of key usability metrics between pairs of 
interfaces. Statistically significant pairwise differences are reported in the 
text.
}
\label{tab:pairwise-contrasts}
\end{table*}

Building on our qualitative findings in RQ1, we saw that participants' use of \systemName{}'s features made them more confident in their ability to find information important to them in the papers. This support manifested differently based on the \systemName{} features available to a participant. There were two major ways we saw \systemName{} improving participants' reading experience: providing in-situ support with the \plainSum{} and \defs{} and a high-level overview with the \keyQs{}.     

The in-situ nature of the \plainSum{} and \defs{} helped participants understand the paper without switching contexts  (P2, 6, 7, 11, 16--17, 19). For example, P19 found the Term Definitions useful for understanding the paper and the more specific medications it mentioned. P2 reported that the \plainSum{} were helpful to understand the paper text in a language they understood and P17 found the \plainSum{} broke ``down complicated medical text into layman's terms that are easily understandable and helped to keep up with the flow of the article.'' While participants could search for definitions of terms and potentially make sense of a passage with a search engine, both activities require turning away from the paper itself. This context switching can make it difficult to keep a thread of reading, especially when that reading is demanding. Our observations suggest that \systemName{}'s in-situ support successfully provided information to participants with minimal context switching. 


Participants also used the \keyQs{} to get an overview of a paper quickly and easily, boosting their confidence to then dive in to the paper text (P2--3, 9--11, 20). P9 reported that ``with so many sample sizes, numbers, and information to go through, it was helpful to get a summary to direct my reading and understanding.'' P20 mirrored this sentiment, explaining that the simplified answers gave them the gist of the entire paper quickly, so they had more time to get into its details. P3 illustrated these benefits well, explaining that the \keyQs{} were ``beneficial because\ldots{}I could have a baseline of what to expect and my mind would not have to pull in many random parts of information and could easily block what I did not need when I only needed a couple bits while I was reading.'' Similar to how the \keyQs{} supported a non-linear reading strategy (described in \S\ref{sec:RQ1}), it seemed that the \keyQs{} allowed participants to get a general sense of a paper early and focus their reading to sections they found most important.  


The \keyQs{} provided useful guidance for readers in this reading context. As shown in Table~\ref{tab:pairwise-contrasts}, readers that only had the \keyQs{} rated their reading difficulty significantly lower ($\tilde{x}=3.00$, $\sigma=0.97$) than participants with the baseline PDF reader ($\tilde{x}=4.00$, $\sigma=1.04$). Participants with the \keyQs{} also rated their confidence that they got all relevant information in a paper ($\tilde{x}=4.00$, $\sigma=0.94$) and that they understood the paper ($\tilde{x}=4.00$, $\sigma=0.89$) significantly higher compared to the PDF baseline ($\tilde{x}=2.50$, $\sigma=1.00$ $\tilde{x}=2.00$, $\sigma=1.00$). 


 
The preference for the \keyQs{} illustrates the importance of the novel guidance technique in \systemName{}. 18 out of 20 readers who had the \keyQs{} in at least one condition selected the Key Question index, not the Answer Gists, as the most helpful feature. P18, who selected the Key Question index as the most helpful feature, said they would absolutely use the questions, because ``...medical papers are difficult to follow and understand without guidance.'' Participants reported liking the Key Questions for quickly finding and understanding relevant information (P2, 4, 7-10, 13, 18-20). P4 reported not having any idea how to approach the research papers, and the Key Questions helped guide them to questions they should have. P7 used the Key Questions because ``It answered questions that I would have had if it was me in the scenario \ldots{} it helped highlight directly to the passage instead of having to sift through all of the information.'' These findings support the insight of this paper that novel guidance-offering features are important for supporting readers in approaching medical research papers.







\subsection{Do we observe any difference in paper comprehension when participants use \systemName{}?}

\begin{figure}[]
    \centering
    \includegraphics[width=0.6\textwidth]{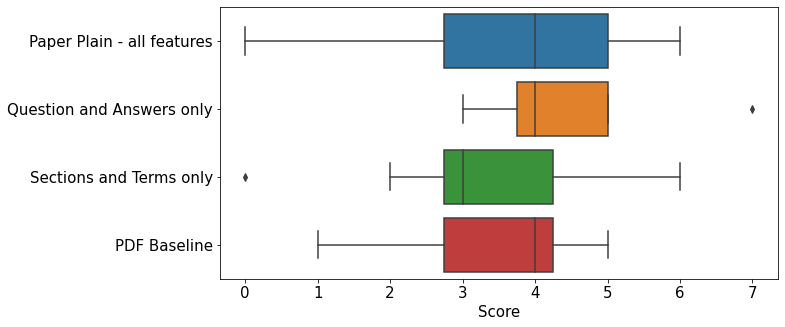}
    \caption{Comprehension scores for variants of \systemName{}. Score is number of comprehension questions answered correctly out of seven for each paper.}
    \label{fig:comprehensionScores}
\end{figure}

Participants on average answered $3.73$ ($\sigma=1.51$) out of $7$ comprehension questions correct. This is well above a random chance baseline of $1.75$ (25\% of 7--each question had four equally likely answers), suggesting that the comprehension questions were answerable given adequate paper reading. Participants scored no worse on the comprehension questions with \systemName{} ($\mu=3.67, \sigma=1.78$) compared to the PDF reader ($\mu=3.50, \sigma=1.31$) (Non inferiority t-test $t_{28}=1.82$, $p < 0.05$). Figure~\ref{fig:comprehensionScores} plots the comprehension scores for each system variant. 


The scores suggest that \systemName{}'s primary objective of improve reading experience was achieved while not hindering comprehension. We designed the comprehension questions so that their answers could be found in the original paper text, not in the gists.  For this reason, it is not surprising that participants scored similarly on the comprehension questions across variants. While comprehension was not the primary objective of \systemName{}, improving readers' understanding of medical papers is also important for ensuring they have productive conversations with their healthcare providers and make informed decisions about their health. We discuss future interventions focused on improving paper comprehension in \S\ref{sec:futureWork}. 

Participants generally found the generated gists useful, and when confronted with vague system predictions and generations, participants usually, though not always, used the original text to fill in missing information. We observed one participant, P11, who read only the Answer Gists for a paper and rated their confidence for understanding the paper at a 5 (the highest) while rating their reading difficulty at a 1 (the easiest). However, this participant got only 2 out of 7 comprehension questions correct, well below the average of $3.73$ for all participants, suggesting that the gists were not sufficient for answering many of the comprehension questions. In contrast to this participant, other participants reported that the gists (both Answer and Section) were helpful as a starting point for understanding, but looked at the underlying text, too. Some participants also reported that information in the gists was vague or missed information in the original text, necessitating reading the original (P10, 22, 24). P24 made sure to double check all the information in the gists with the original sections because the gists were automatically generated. While they did not find incorrect information in the gists, they did report that the Sections Gists sometimes were vague or reported on details less important to them while leaving out details that were more important to them (e.g., the percent of people who recovered from a surgery was reported in a section but not the Section Gist). P10 also noticed that the area surrounding some of the highlighted answers contained useful information, and so made sure to go back through the answering passages to read the surrounding text in addition to the Answer Gist and passage. While it seems that most participants found the gists useful and read the original text alongside the generated gists, we discuss future designs to encourage reading original paper content in \S\ref{sec:designImplications}.

\section{Discussion \& Future Work}
    
This paper illustrates how interactive information interfaces can make research papers approachable to healthcare consumers that need it. In particular, we develop \systemName{}, an interactive system that augments the paper itself with new affordances to help readers navigate, evaluate, and understand its contents. 


\subsection{Summary of the results}

\textit{How did participants use \systemName{}'s features?} Participants used and appreciated \systemName{}'s features throughout reading a paper. Readers used the \plainSum{} to easily make sense of dense passages while reading a paper and leveraged the guidance of the \keyQs{} to quickly find text that was informative for them. All but one participant said they would use \systemName{} to read medical papers.

The \keyQs{} were a clear favorite in the usability study. When participants had access to all features, they used the \keyQs{} more often than the \plainSum{} and \defs{}. Participants used the Key Questions to jump to sections informative to them compared to participants with the typical PDF reader or the \plainSum{} and \defs{}. These results suggest that readers took advantage of the questions' affordance of fast-tracking to the important information in a paper.


\textit{How does \systemName{} affect participants' self-reported reading difficulty, understanding, and ability to identify relevant information?} Participants who used \systemName{} rated their reading difficulty significantly lower and rated their confidence they got all relevant information from a paper significantly higher. Participants found it easier to read with \systemName{} because it gave them an approachable overview of a paper with the \keyQs{} and helped them understand dense text in the context of the paper with the in-situ \defs{} and \plainSum{}. 

It is worth noting that non-experts can be overly confident in their understanding of scientific material~\citep{Scharrer2017WhenSB} and therefore our subjective ratings of understanding should be judged with caution. That being said, the strong results for reading difficulty suggest that \systemName{} was able to support readers in overcoming some of the barriers we observed for reading medical research papers.  




\textit{Do we observe any difference in paper comprehension when participants use \systemName{}?} Participants who used \systemName{} had similar comprehension scores compared to participants using the typical PDF reader. While improving comprehension would have been an exciting addition to our findings that \systemName{} improved reading experience, the similar scores in comprehension provide compelling evidence that \systemName{} achieves its primary focus of lowering barriers to paper reading without any loss in paper comprehension.






In summary, we take these results to indicate the promise of \systemName{} for assisting healthcare consumers in making sense of medical research papers. The sustained usage of \systemName{}'s features and positive response from participants in our usability study suggest that such a tool would be a welcome addition to healthcare consumers' information seeking toolkit.

\subsection{Design implications}\label{sec:designImplications}



This paper’s exploration of features can inform future interactive reading systems. We offer the following guidance for developing such future systems: 

\textit{Provide reading guidance} Interactive reading systems for non-experts can provide more active support for guiding readers. Experts have strategies to quickly gather relevant information in a paper without engaging in a deep read (e.g., skimming)~\cite{Shanahan2011AnalysisOE}, but most readers in our studies didn't have a particular strategy for reading the papers, defaulting to an exhaustive linear pattern. This led to readers getting stuck in dense passages with minimal relevance to their scenario. 

The features in \systemName{} offering guidance (i.e., the \keyQs{}) led to the largest improvements in reading experience by providing an alternative reading strategy. Readers with the Key Questions in our usability study jumped to important sections of the paper within the first few minutes of reading. While we did observe some tension in our iterative design where the Key Questions distracted readers who wanted to approach papers on their own, it seems that our final design for the feature (a toggleable sidebar) offered useful guidance without distracting readers. Indeed, participant feedback supports this: 18 out of 20 participants selected the Key Questions themselves, not the Answer Gists, as the most helpful feature. 

\textit{Incorporate plain language into the original document} While guidance was helpful for directing readers' effort to sections of interest, readers still needed plain language to lower the effort needed to understand those sections. Every participant in our usability study with access to plain language features, either Answer or Section Gists, used them to make sense of the papers. 


At the same time, any plain language should be in service of making the original document easier to read, not replacing it. Plain language from current generative models can contain inconsistencies~\cite{Maynez2020OnFA}, which risks misinforming readers. \systemName{} encourages readers to focus on the original paper text by having readers pull gist content rather than displaying it immediately with the paper and by placing gist content alongside the paper. Future systems could go further by reporting factuality measures along with generated text~\cite{Pagnoni2021UnderstandingFI} or integrating a feedback mechanism for reporting inconsistent information in order to crowd-source these factuality checks.

\subsection{Ethical and Social Implications}\label{sec:ethicalImplications}

While we believe that lowering barriers to reading medical research papers can benefit healthcare consumers by informing them about their care, there are certain risks as well. 
One issue is that healthcare consumers can be unaccustomed to norms in the scientific process, such as the fact that a single paper does not represent scientific consensus. Readers might mistake findings or interpretations in a paper as truth, which could risk them making misinformed decisions about their care. At the same time, readers are already taking these risks and turning to medical research papers~\cite{Day2020OpenTT}, and \systemName{} can help them understand a paper more easily than if they were on their own. 

A key limitation of current generative models is their propensity to hallucinate, generating factually inconsistent or incorrect information~\cite{Maynez2020OnFA}. These hallucinations could misinform readers, which would be extremely costly in the context of personal health information seeking. There is growing interest in evaluating factuality in generations (for common factuality measures, see ~\cite{Gabriel2021GOFA, Fabbri2021SummEvalRS, Pagnoni2021UnderstandingFI}). We are excited to integrate new advances for measuring and ensuring factuality in generations (e.g., \cite{Laban2021SummaCRN, Dong2020MultiFactCI, Nan2021ImprovingFC}) into \systemName{} to help realize the promise of such models in real-world settings.

\subsection{Limitations}
\label{sec:limitations}


Recruiting participants on Upwork might have have skewed our barriers and resulting design since participants were not reading medical papers that were personally relevant to them. This could have led participants to pay less attention to specific paper details or get less discouraged by negative findings and unclear results. We designed the studies to be as close as possible to the real work healthcare consumers reading medical research papers engage in by writing scenarios based on our findings from interviews with such readers. 


The usability study was also a timed, relatively short (10 minute) reading task, which might have skewed participant reading habits. It could be that differences in comprehension would become more pronounced if more time was given. Additionally, patterns of usage of the interface may start to look different after the first 10 minutes of reading. Some participants reported that if they had more time, they would have read the paper through again or looked for additional information. Others felt that the time limit made them anxious and they had a hard time remembering information. This might have also artificially inflated participants' use of the \keyQs{}, since they offered the fastest way of getting an overview of the paper. Participants also might have used the \plainSum{} more if given more time since the \plainSum{} were helpful at allowing participants to go off on their own to explore sections of the paper. Because reading time could influence comprehension and subjective reading confidence, we decided it was important to keep time consistent across the study. In future work we are excited to observe the use of \systemName{} in more naturalistic settings.

\subsection{Future directions}
\label{sec:futureWork}


Our studies and system, \systemName{}, reveal exciting areas of future research in information interfaces for medical information-seeking and augmented reading interfaces broadly. We discuss a few of these directions below. 


\textit{Enabling intelligent reading interfaces} As AI technology advances, new interfaces integrating this technology can provide tremendous value to users. Our work illustrates a path towards one such system with \systemName{}. When developing \systemName{}, we mapped its features to existing natural language processing (NLP) techniques like biomedical question answering (QA)~\cite{10.1007/978-3-030-43887-6_64} and plain language summarization~\cite{Brown2020LanguageMA}. There are additional NLP techniques that could augment reading experiences, such as machine translation~\cite{Kirchhoff2011ApplicationOS}, toxic language detection~\cite{MacAvaney2019HateSD, Hosseini2017DeceivingGP} or news story mapping~\cite{Laban2020AFF}. We hope that our discussion in \S\ref{sec:implementation} of techniques to make machine output useful for readers (e.g., by providing full paragraphs rather than a single word in QA output) can provide useful insight for future reading interfaces integrating machine intelligence. 

\textit{Improving paper comprehension}  While our results show that \systemName{} improved reading difficulty and confidence by addressing the barriers revealed in \S\ref{sec:barriers} without any loss in comprehension, an important future step is to design interventions for explicitly improving paper comprehension. Simplifying scientific information can risk over-inflating readers' sense of understanding and reduce their reliance on experts, even when such judgements are misplaced~\cite{Scharrer2017WhenSB}. One possible way of improving comprehension would be to focus on protecting against common misunderstandings for healthcare consumers reading medical literature~\cite{Day2020OpenTT}, such as by identifying predatory journals without peer review or by providing findings summarized from multiple papers.

\textit{Supporting healthcare providers and patient advocates} Our work focused on making medical research papers more understandable to healthcare consumers, specifically patients and caregivers, but \systemName{} could also benefit other stakeholders in medical research. For example, healthcare providers and patient advocates often need to read medical research papers to apply the findings in clinical practice~\cite{Rabeharisoa2013EvidencebasedAP, Guyatt1992EvidencebasedMA, Brownson2018BuildingCF}. The information needs and barriers of these groups differ from healthcare consumers, necessitating extensions of \systemName{}. Providers have to handle many more research papers covering many patients. For these needs \systemName{} could extend to cover multiple papers at a time, such as by extracting answers for key questions across the papers. At the same time, providers can draw upon their medical experience for understanding papers, so \systemName{} could focus on extraction and summarization over plain language. 


\textit{Addressing additional barriers for healthcare consumers} Extensions of \systemName{} could relate information in papers to readers' personal circumstances. During our formative study observing readers (\S\ref{sec:barriers}), some participants expressed interest in knowing patient testimonials about treatments in the paper or wanting to know how individual patients most similar to the reader responded to treatments. While the current design of \systemName{} did not address this need since it required information not available in the paper (e.g., patient testimonials), it is exciting to imagine future systems that draw on other information sources, such as online support groups, to relate the information in medical papers to reader personal experience.



\textit{Supporting non-experts in other domains} Medical research is not the only context where non-expert readers wrestle with highly technical documents; \systemName{}'s design can inspire efforts in addressing similar barriers in these other contexts. Some aspects of these contexts merit new design efforts, while others might benefit from similar designs as \systemName{}. For example, the important questions to ask while reading a medical paper are different than those for a legal contract or privacy statement. This necessitates a re-crafting of the \keyQs{} for other domains. Furthermore, some documents may need to be read in a particular order (e.g., a software tutorial), and providing an alternative index, as \systemName{} does, could confuse readers. In these cases, any key question indexes into the document would need to be aligned with the document's original structure. Providing in-situ \plainSum{} and \defs{} could help address needs in these domains, as these features can help readers understand what they are reading within the flow of a document.

\section{Conclusion}

In this paper we ask how interactive interfaces can make medical research papers approachable to healthcare consumers that need it. Our key insight is that medical papers can be made more approachable by incorporating plain language summaries alongside original paper content and providing guidance on the most important passages to read. We illustrate these interactions with a novel system, \systemName{}, which draws on active research in natural language processing to show the potential for automated support in this reading context. In a usability study of \systemName{}, we found that participants who use \systemName{} have an easier time reading research papers compared to those who use a typical PDF reader, and that participants most appreciated the reading guidance offered by \systemName{} via a key question index. While further algorithmic advances are required to ensure a safe deployment, we envision \systemName{} as a system that can one day be enabled for any medical research paper.

\begin{acks}
This work would not have been possible without the contributions of others. Joseph Chang, Hyeonsu Kang, Dan Weld and Amy Zhang provided feedback on the paper. Matt Latzke taught us how to use Figma for iterative design. Sam Skjonsberg provided mentorship during the development of the prototype UI. Bailey Kuehl assisted us with recruiting study participants on Upwork. Jay Bartot and Keshet Ronen advised us on the perspectives of healthcare consumers, caregivers, and providers. Katharina Reinecke and Noah Smith provided general feedback on the project. This research was performed during an internship at the Allen Institute for AI with the Semantic Scholar team.

\end{acks}

\bibliographystyle{ACM-Reference-Format}
\bibliography{sample-base}

\pagebreak{}
\appendix
\section{Interviews with healthcare consumers and providers}
\label{app:motivations}

To validate the general idea of helping readers understand medical research papers, we interviewed six healthcare consumers and providers. We spoke with both healthcare consumers with prior experience reading medical research (4 total, referred to as C1--4), and healthcare providers who had discussed findings from medical papers with their patients (2 total, H1--2). Healthcare consumers and providers were recruited through our personal and professional networks and by referral from other interviewees.

These interviews yielded a set of scenarios in which readers turn to the medical literature. These scenarios motivated the design of our interface and are offered here to inspire future research to help readers engage with the medical literature.

Our participants read medical literature because they wanted more information than they could gather from discussions with their doctor or by consulting conventional patient-facing resources online. This core motivation manifested in four cases:

\begin{itemize}
    \item \textbf{Learning more about the diagnosis}: Participants’ expressed a desire to know more information than what patient pamphlets or their short doctors’ appointments could give them because they wanted to understand the diagnosis in greater depth (C1, C3). 
    \item \textbf{Learning background-specific information} Participants sought the medical literature because their situation was somewhat unique compared to the common diagnosis (e.g., affecting a different part of their body or at a different age) (C1, C2).
    \item \textbf{Becoming aware of emerging treatment options}: Participants mentioned that having chronic illnesses or those without cures (e.g., severe allergies), had encouraged them to seek out new clinical trials and trial results as a way of finding new ways of taking care of themselves. (C1, C4)
    \item \textbf{Comparing treatment options}: Similar to finding new treatments, participants described trying to decide between different treatments their doctor recommended or just wanting to know more about these treatments (e.g., results from clinical trials or alternative treatments) (C1). 
\end{itemize}

These findings support prior work on motivations in consumer health information seeking
~\cite{Sommerhalder2009InternetIA} and illustrate the benefits of open-access medical literature~\cite{Zuccala2010OpenAA} as an additional resource for healthcare consumers to find information important to them.  A healthcare provider we spoke to gave similar insights: their patients sought medical research papers as a source of information to supplement in-person discussions with their physician (H1).

Conversations with our participants suggested that paper reading presented issues such as unfamiliar terminology, assessing relevance, and information overload. C1 and C3 mentioned that many paper titles were already too complex, or they needed to learn a lot of new medical vocabulary as they read. C4 described the emotional exhaustion of reading through multiple discouraging results. C2 mentioned how hard it was to assess if research was trustworthy or relevant to them. All participants mentioned only being able to engage with research papers for an hour or two before they were exhausted. To develop a deeper understanding of how these challenges manifest during reading, we designed a second formative study where we observed non-experts as they encountered these challenges when reading medical papers.

\section{Iterative Design}\label{app:iterative_design}

A total of eight participants (N1-8) used two early prototypes of \systemName{} in qualitative usability evaluations. Participants were recruited from our institution, our professional networks, and Upwork. In these evaluations, participants were given a modified scenario from \S\ref{sec:barriersProcedure} and read a paper with the \systemName{} prototype. These evaluations lasted one hour each.






Overall participants reported that using the \systemName{} prototypes helped them access important information in a paper (N1--6, 8). Participants said that the features helped them focus their attention while reading (N4) and gave them a good overview of the paper (N1 and 3). Participants all expressed excitement for such a tool existing for their own health information seeking. The usability evaluations also illustrated important design goals for effective interactive aids in this reading context, which we integrated into the design of \systemName{}:

\textbf{Provide gists on-demand.}
Plain language is not just useful for helping readers understand the text; it can also help readers avoid reading an abundance of dense text. Providing plain language throughout a paper can help readers choose what not to read. N1 used a prototype with only plain language answering passages (``Answer gists'') and reported that having only answering passages simplified was restricting their ability to explore the paper on their own. N3 also wanted gists for scanning other sections of the paper that might not have an answering passage, such as specific results sections. 


\textbf{Make guidance both discoverable and unobtrusive.} Readers often don't know where to look for relevant information in research papers. Navigation that guides readers to relevant sections can save them time and effort, even if it reduces some of their autonomy. 

The Key Question Index gave an accessible overview of a paper, but participants often did not notice the sidebar toggle until they had spent considerable effort understanding the paper. For example, two participants (N1 and N3) missed the button to toggle the Key Question Index sidebar, and only noticed it later in the session when it was pointed out by a researcher. After seeing the Key Question Index, N1 mentioned that they wished they had seen it earlier since it would have provided a helpful high level understanding early on. 


At the same time, the sidebar could be intrusive to some participants. One participant (N5) reported that the sidebar was distracting and occluded other typical PDF reader features they wanted to access, such as section outlines. To balance the goal of providing an intuitive guide without clashing with readers' other potential reading strategies, \systemName{}’s final Key Question Index sidebar was opened when a paper loads but was toggleable to other sidebars and able to be closed.



\textbf{Supplement, rather than replace, the text}. The text is critical; it is where readers will find nuanced details that would not be available in summaries or conventional healthcare consumer materials. Features should make the text more understandable, not replace it. In addition, NLP systems are imperfect, and a reader who relies solely on generated content can risk misunderstanding the actual paper. N1 often double-checked gists with the original text and N4 hid the gists to read the underlying text. We wanted to make sure that the system focused readers on the original text and provided generated text as a supplement, not a replacement. In the prototype the gists were sometimes overlapping the original text, which made it hard for participants to read both. In the final design of \systemName{}, all gists were placed as close to the original text as possible without overlapping. Furthermore, gist content was provided on-demand, rather than initially displayed along with the paper, to encourage readers to focus on the paper and pull supplemental content from the gists only when necessary.  


\section{\systemName{} Implementation}
\label{app:implementation}

\subsection{GPT-3 Simplification}

We adapted our GPT-3 prompt and generation parameters (e.g., length of generation and temperature) from one of the preset examples that OpenAI provides for summarizing text for a 2nd grader.\footnote{\url{https://beta.openai.com/examples/default-summarize}} We changed the prompt to summarize for a 5th grader rather than 2nd grader after observing that using 2nd grader caused the model output to be too general and vague. We also tested later grades, up to college, but found that the generated text using the 5th grader prompt was the most consistent.  Our final prompt for GPT-3 was: 


\begin{quote}
My fifth grader asked me what this passage means:
"""
[TEXT TO SIMPLIFY]
"""
I rephrased it for him, in plain language a fifth grader can understand:
\end{quote}

\noindent We also updated generation parameters, specifically the length of generation and temperature (a parameter for controlling the randomness of generations). We set generation length to 100 characters and temperature to a range of $0.25$ to $0.5$, depending on the generation.





\section{Statistical Analysis}
\label{app:stats}

\subsection{Modeling Mixed-Effects in Repeated Measures Studies}

For the analysis in \S~\ref{sec:usabilityAnalysis}, we used the linear 
mixed-effects model (LMM). LMMs are commonly used to analyze data in which the same participant provides multiple, possibly correlated, measurements, referred to as repeated measures~\cite{Lindstrom1990NonlinearME}. LMMs are used as an analysis in medicine~\cite{Cnaan1997UsingTG}, the behavioral sciences~\cite{Cudeck1996MixedeffectsMI}, and human-computer interaction~\cite{Hearst2020AnEO, Head2021AugmentingSP}.

For each of the quantitative measurements discussed in \S\ref{sec:measures} ($y$), we fit a LMM with fixed 
effects $\beta$ for the \systemName{} paper ($x_1$) and interface variant ($x_2$) factors.\footnote{We also fit the same LMM with an additional interaction term ($x_1 x_2$) but the $F$-test for this term was not significant across the three measures ($p > 0.67$, $p > 0.98$, $p > 0.98$). As such, we proceeded with our analysis without the interaction term in our LMM.} We used the \textsc{lme4} 
package in R~\cite{Bates2014FittingLM} to fit the models. More precisely, we fit the following 
LMM:

\begin{equation}
    E[y] = \beta_0 + \gamma_j + \beta_1 x_1 + \beta_2 x_2,
\end{equation}

\noindent where the random intercepts $\gamma_j \sim 
\mathcal{N}(0, \sigma^2_{\gamma})$ capture individual variation of each 
participant $j$. 

We report all the estimated coefficients in Table~\ref{tab:lmm}. Due to the categorical nature of our variables, we interpret the coefficients in the following way: $\beta_0$ is the mean score for \systemName{} while reading the paper for herniated disc. $\beta_1^{SLE}$ is the mean difference in score for the SLE paper, given the same interface variant. Similarly, $\beta_2^{PDF}$, $\beta_2^{SD}$ and $\beta_2^{QA}$ are the mean differences in score for the PDF baseline, \defs and \keyQs{} interface variants against full \systemName{} variant, given the same paper. For example, $\beta_2^{PDF} = 1.9835$ for Reading Difficulty means that the PDF baseline is associated with a 1.9835 higher difficulty score than \systemName{}, which is the same result we report in Table~\ref{tab:pairwise-contrasts}.




\begin{table*}[h!]
\centering
\begin{tabular}{rccccc}
\toprule 
&
$\beta_0$ &%
$\beta_1^{SLE}$ &%
$\beta_2^{PDF}$ &%
$\beta_2^{SD}$ &%
$\beta_2^{QA}$ 
\\

\midrule              

Reading Difficulty (1--5)&%
2.0884 &%
0.3750 & %
1.9835 &%
1.4851 &%
0.3444 %
\\

Understand (1--5) &%
3.8231&%
-0.5000 & %
-1.1769 &%
-0.7194 &%
0.1037 %
\\

Relevance (1--5)&%
3.9316 &%
-0.5833 & %
-1.1675 &%
-0.7524 &%
0.1934 %
\\

\bottomrule

\bottomrule
\vspace{0.2ex}

\end{tabular}

\caption{
Estimated fixed-effect coefficients for the LMM described in Appendix~\ref{app:stats} for each measurement.
}
\Description{%
}
\label{tab:lmm}
\end{table*}

\subsection{F-Tests for Significant Effect of Interface}



We conducted \emph{F}-tests for differences in fixed-effect estimates between each 
interface variant, repeated for each $y$ using the \textsc{lmerTest} R package~\cite{Kuznetsova2017lmerTestPT}. Using the Holm-Bonferroni \cite{Holm1979ASS} correction on the $p$-values with the \textsc{p.adjust} R package, we found
significance for reading difficulty $(p<.001)$, relevance $(p<.001)$, and confidence $(p<.001)$---even while 
controlling for paper and participant-specific effects. That is to say, for 
these metrics, the \emph{F}-test identified that the choice of interface 
(\systemName{}, \navGuide{}, \inSituExpl{}, or \pdf{}) is a significant factor.  Note that the 
\emph{F}-test does not identify \emph{which} interfaces differ from one another on the metric.  

\subsection{Tests for Pairwise Differences in Fixed-Effects between Interfaces}

To quantify the pairwise differences in fixed-effects between the interface variants for the measures $y$ under the LMM (and controlling for paper), we conducted a post-hoc analysis.  We used two-sided $t$-tests for pairwise comparisons using the 
\textsc{emmeans} R package, yielding the results shown in 
Table~\ref{tab:pairwise-contrasts}.


\subsection{Ordinal Regression for Likert-Scale Variables}

As reading difficulty, confidence, and understanding were measured on a Likert scale, a LMM 
estimated means could be ill-suited for analysis, especially 
if these measures were not sufficiently normally distributed. We additionally 
performed likelihood ratio tests after fitting analogous cumulative link 
mixed-effects models (CLMM) provided in the \textsc{ordinal} R package 
\cite{Christensen2018CumulativeLM}. Likelihood ratio tests, which are similar to 
\emph{F}-tests but more conservative, yielded similar $p$-values---reading difficulty ($p < 
.001$), confidence ($p < .001$), and understanding ($p < .001$) ---and resulted in the same conclusions as those 
when using the LMM.  Because pairwise analyses were not available through 
\textsc{emmeans} (or other libraries) for CLMMs, we opted to use the LMM model 
for these measures to enable subsequent analysis for Table 
\ref{tab:pairwise-contrasts}.

\begin{table*}[t]
\centering 
\begin{tabular}
{L{13mm}L{30mm}L{45mm}L{45mm}}
\toprule
 \textbf{Source} & \textbf{Question} & \textbf{Extracted Answer}  & \textbf{Plain Language Answer}     \\ \midrule \midrule 
    PICO & What condition does this paper study? & ``Systemic lupus erythematosus (SLE) is the prototypical auto-immune connective tissue disease\ldots{}'' &  ``Systemic Lupus Erythematosus is a disease that affects about 5 million people in the world\ldots{}'' \\ \\
    PICO & How is the condition usually treated? & ``Following the diagnosis of SLE, patients are assessed for disease activity and organ involvement, both of which dictate the most appropriate therapy\ldots{}'' & ``After you get the diagnosis of lupus, the doctor will see how bad your lupus is and how much it affects your body\ldots{}'' \\ \\
    Cochrane & What did the paper want to find out? & ``The aim of this review is to report the evidence concerning the rationale, the efficacy, and the safety of therapeutic peptides\ldots{}''  & ``This is a review of the evidence and reasons why doctors are using peptides to treat lupus\ldots{}'' \\ \\
    Cochrane & What did the paper do? & ``The next paragraphs report and discuss the current evidence concerning unconjugated and conjugated therapeutic peptides\ldots{}'' & ``The next paragraphs tell us about some drugs that are being tested to see if they can help people with lupus\ldots{}'' \\ \\
    PICO & What were the new treatment(s), if any this paper looked into? & ``Therapeutic peptides include a class of pharmaceutical com- pounds consisting of amino acid chains of various length (usually less than 40 amino acids)\ldots{}'' & ``A peptide is a small molecule made up of amino acids, which are the building blocks of proteins\ldots{}'' \\ \\
    Cochrane & What did the paper find? & ``To date, no therapeutic peptide has been licensed and marketed for the use in SLE patients\ldots{}''  & ``A drug that targets a specific part of the immune system is being tested to see if it can help people with a disease called lupus\ldots{}'' \\ \\
    PICO & Are the findings different depending on a person's demographics? & ``Being designed on the basis of epitopes that are pathogenic in SLE alone, peptides\ldots{}'' &  ``These new drugs are designed to target the bad proteins that cause SLE\ldots{}''\\ \\
    Cochrane & What are the limitations of the findings? &  ``Nevertheless, despite the successful results observed in preclinical studies, RCTs showed a controversial efficacy profile\ldots{}'' & ``Even though the medicine worked well in the lab, it did not work as well in real life\ldots{}'' \\ \\

\bottomrule

\end{tabular}
\caption{Questions used in \keyQs{}. Questions are presented in the order they appear in the index.}\label{tab:questions}
\end{table*}

\end{document}